\documentclass[12pt]{article}
\usepackage{geometry}
\usepackage{a4}
\usepackage{graphicx,subfigure}
\usepackage{epsf}
\usepackage{amsmath}
\usepackage{amssymb}
\usepackage{cite}
\newcommand{\be}{\begin{equation}}
\newcommand{\ee}{\end{equation}}
 
 \newcommand{\Rmnum}[1]{\expandafter\@slowromancap\romannumeral #1@}
\newcommand{\bea}{\begin{eqnarray}}
\newcommand{\eea}{\end{eqnarray}}

\begin{document}
\def\A{{\mathbb{A}}}
\def\C{{\mathbb{C}}}
\def\R{{\mathbb{R}}}
\def\s{{\mathbb{S}}}
\def\T{{\mathbb{T}}}
\def\Z{{\mathbb{Z}}}
\def\W{{\mathbb{W}}}
\begin{titlepage}
\title{Holographic Thermalization with Weyl Corrections}
\author{}
\date{
Anshuman Dey, Subhash Mahapatra\footnote{Present Address : The Institute of Mathematical Sciences, Chennai 600113, India.}, 
Tapobrata Sarkar
\thanks{\noindent E-mail:~ deyanshu@iitk.ac.in, subhmaha@imsc.res.in, tapo @iitk.ac.in}
\vskip0.4cm
{\sl Department of Physics, \\
Indian Institute of Technology,\\
Kanpur 208016, \\
India}}
\maketitle\abstract{
\noindent
We consider holographic thermalization in the presence of a Weyl correction in five dimensional AdS space. 
We first obtain the Weyl corrected black brane solution perturbatively, up to first order in the coupling. The corresponding 
AdS-Vaidya like solution is then constructed. This is then used to 
numerically analyze the time dependence of the two point correlation functions and the expectation values
of rectangular Wilson loops in the boundary field theory, and we discuss how the Weyl correction can modify the thermalization
time scales in the dual field theory.  In this context, the subtle interplay between the Weyl coupling constant 
and the chemical potential is studied in detail. }
\end{titlepage}

\section{Introduction}
The AdS/CFT correspondence \cite{Maldacena}, \cite{Klebanov}, \cite{Witten} or the gauge/gravity duality is one of the most 
striking aspects of string theory and is being extensively used over the past few years to study strongly coupled condensed
matter systems. The duality relates a classical theory of gravity in a $(d+1)$ dimensional anti-de Sitter (AdS) spacetime to a 
strongly coupled conformal field theory living on the boundary of the AdS space in $d$ spacetime dimensions. The strongly coupled
nature of the boundary theory does not allow the usage of standard perturbative techniques.
However, using the gauge/gravity duality, the computation becomes simpler to handle, because the dual gravity theory is classical. 
This is the primary motivation to explore phenomena in strongly coupled quantum systems
from a holographic point of view. It is probably fair to say that by now, a clear understanding of the near-equilibrium physics
of strongly coupled quantum field theories arising from the dual gravity sector has emerged. For example, one can calculate the bulk 
correlation functions \cite{Son} and compute different observables holographically and understand the linear response of the system to
perturbations from equilibrium \cite{Son1}, \cite{Hubeny}. However, it is quite difficult to 
understand the physics of a strongly coupled system, even from the dual gravity sector, when the system is out of equilibrium,
because now one can not apply linear response theory. It is in fact very interesting to analyze how such a  
system reaches thermal equilibrium, once it is out of equilibrium, and calculate the ``thermalization time.'' In a class of examples, 
this issue has been resolved holographically by constructing a time-dependent gravity solution in AdS space which describes the 
formation of a black hole at late times. 

The other fact that motivates the study of non-equilibrium dynamics from the gauge/gravity duality, is the experimental
input from the Relativistic Heavy Ion Collider (RHIC) and the Large Hadron Collider (LHC). When two large energy heavy 
ions collide in RHIC, some of their kinetic energy is transformed into heat energy. Because of the large amount of heat produced,
the quarks and the gluons form a plasma-like state, known as quark gluon plasma (QGP). The process of forming QGP
is known as thermalization. After the QGP is formed, i.e., after the thermalization process is over, it reaches the 
thermal equilibrium where one can apply linear response theory to understand the physics of QGP. The experimental result shows that the 
QGP formed in RHIC behaves like an ideal fluid, with a very small shear viscosity to entropy density ratio ($\eta/s$) indicating the strong 
coupling nature of the QGP \cite{Shuryak}. Because of the strong coupling constant it is appropriate to use the
AdS/CFT correspondence to compute $\eta/s$, and check whether they match with the experimental results.
A lot of work \cite{Son1}, \cite{Policastro}, \cite{Policastro1}, \cite{Buchel}, \cite{Brigante} has been done in this
direction concluding that there exists a small lower bound of the ratio which may depend on the coupling constant of the
theory. While it is easy to compute different observables after the QGP is formed, it is difficult to compute the 
thermalization time, since, as mentioned before, thermalization is a non-equilibrium process. Also the observed thermalization time in RHIC is
much shorter than what is predicted by calculations via perturbative techniques \cite{Heinz}. This indicates that
the thermalization process also takes place within a strong coupling regime of QCD. These experimental results provide further important
motivation to analyze thermalization, using the gauge/gravity duality. We also point out that there are large number of articles 
\cite{Danielsson}-\cite{Caceres} where the thermalization process has been described as the dual of a black hole formation by a gravitational
collapse in the bulk. 

In \cite{Bala} and \cite{Bala1}, the authors considered an interesting model for thermalization in the context of AdS/CFT, 
and examined the thermalization after a sudden injection of energy to the boundary field theory by  three different non-local probes : 
two-point correlator, Wilson loop and entanglement entropy.
All of these three probes are well-defined in the dual gravity theory,  and they are described by different geometric quantities.
For example, the two-point correlator on the boundary corresponds to a geodesic connecting the two points and extending 
into the bulk. Similarly, the expectation value of the Wilson-loop operator and the entanglement entropy correspond to a 
minimal area surface and minimal volume, respectively, extending into the bulk. The model they considered is known as the
AdS-Vaidya metric which describes the collapse of a thin shell of matter from the boundary to the bulk. As the shell 
collapses, it divides the spacetime into two region: the outer region of the shell represents a black brane while the inner
region corresponds to pure AdS spacetime. Hence, at the early time, the AdS-Vaidya metric corresponds to a pure AdS space
(representing a vacuum state of the boundary QFT), while it represents a black brane metric (representing a thermal state
of the boundary QFT) at late times after the shell collapses. 

For all kinds of probes, it was shown in \cite{Bala}, \cite{Bala1} that the UV degrees of
freedom thermalize first and the IR degrees of freedom thermalize later, i.e., the thermalization process is top-down. 
\footnote{In \cite{Cardy}, it was shown that UV modes of a $1+1$ dimensional CFT thermalize faster than the IR modes irrespective of the strength 
of the coupling constant.}
While a standard perturbation technique in QCD predicts that the thermalization process should be bottom-up \cite{Baier}, these papers get an opposite 
behavior. Note that the bottom-up behavior of thermalization in heavy-ion-collisions in a perturbative QCD can be realized 
as follows : when the thermalization initiates, a large number of soft gluons are emitted because of the large collision energy. These soft
gluons collide amongst themselves, and equilibrate quickly to form a thermal bath. Hence, it is the low-energy modes or, the IR modes
which thermalize first. Then the thermal bath absorbs energy from the hard gluons and when the hard gluons lose all their energy, the whole system
thermalizes. However, top-down holographic thermalization is sensible from the dual gravity perspective. Simply put, since the IR modes probe more deeply 
into the bulk, the corresponding thermalization time should also be larger. It was also found that the thermalization time scales with the length
of the probe $l$ as $\tau\sim l/2$. 

The work of \cite{Galante} and \cite{Kundu} extended this model to study the thermalization
in the presence of a chemical potential. These authors modelled the dual gravity theory in such a way that at late time when the shell 
collapses, the corresponding AdS-Vaidya metric would represent a Reissner-Nordstr\"{o}m AdS black brane. They found that
for larger probes, as one increases the ratio of the chemical potential to the temperature, the thermalization time increases. 
Then the idea was generalized to investigate the non-trivial corrections in the thermalization time due to the consideration 
of the higher derivative gravity \cite{Zeng}, \cite{Yang}, Born-Infeld electrodynamics \cite{Abdalla}, Lifshitz and
hyperscaling violating geometries \cite{Reza}, \cite{Tonni}. In a previous article \cite{Johnson}, the authors studied the time evolution of 
holographic entanglement entropy in thermal and electromagnetic quenches using similar kind of model. In \cite{Pedraza}, the author
has studied the late-time behavior of different non-local observables in an expanding boost-invariant plasma and shown how the 
fluid parameters (e.g., shear viscosity) affect the relaxation of these observables. In a recent paper \cite{Reza1},
the time evolution of holographic $n$-partite information has also been investigated.

In a phenomenological i.e., bottom-up approach, it is of substantial interest to understand the process of thermalization  in strongly
coupled field theories upon the inclusion of general four derivative terms in the dual gravity, apart from the leading Maxwell term. 
Such terms are known to give rise to interesting effects - for example they non-trivially affect the $\eta/s$ ratio \cite{Myers}. 
In \cite{Edelstein}, the authors introduced these class of terms in the effective action and showed that they can lead to a violation of
causality which can be prevented by the possibility of pair production. In a top-down
approach, such terms are expected to arise in a string theory as quantum corrections to the low energy effective action. These 
terms are expected to be suppressed in a perturbative sense, and on the CFT side should represent terms suppressed by inverse powers
of the 'tHooft coupling. For example, in the context of five dimensional AdS theories, one can generically think of adding all possible 
four derivative interactions to a usual Einstein-Maxwell action. As explained in \cite{Myers}, and as we review in the next section, 
there are a large number of such terms, but the action can be considerably simplified by choosing particular linear combination of the coupling 
constants. Understanding thermalization in a field theory dual to five dimensional AdS with a generic four derivative action seems a daunting task. 
In particular, the presence of five different coefficients that appear in such a theory is likely to make a general scan of the parameter 
space tedious, since the physics there is likely to depend on (fine tuned) values of the coefficients. 
In this paper, we consider one particular simplified situation. We consider the two derivative Einstein-Maxwell action corrected by a Weyl coupling. 

In the context of holographic thermalization, one of the main motivations for considering such a coupling is the fact that it introduces an 
extra control parameter in the theory which might non-trivially affect the thermalization process. As we have said, a theory with all 
possible higher derivative couplings might be complicated, and in this paper we will see that a Weyl corrected theory already indicates non-trivial
effects, which points to features that might be valid for such a generic theory.  
Indeed, this type of correction has previously appeared in \cite{Ritz}, \cite{Sachdev} who argued that as far as charge transport is concerned, 
starting with a general four derivative term, it is enough to consider only a linear combination of those terms, which involves a coupling of the 
Maxwell field to the bulk Weyl tensor. They computed the correction in the conductivity
and the diffusion constant due to the Weyl coupling constant $\gamma$ and predicted a bound in $\gamma$ from the physical
consistency conditions. This kind of four derivative interaction terms were also encountered before in \cite{Maldacena1} and
\cite{Hofman}. In fact, QED in a general curved background leads to the Weyl coupling term at 1-loop \cite{Drummond}. 

In this paper, we will consider a Weyl correction term along with the two derivative Maxwell term in a five dimensional bulk AdS. Our
purpose here would be to treat this model phenomenologically and compute the effects on thermalization due to the Weyl correction. 
For this purpose, we construct an AdS-Vaidya metric which interpolates between a pure AdS space
at early time  and a black brane solution with Weyl corrections at late time. 
In contrast to the previous works, the AdS-Vaidya spacetime we consider here
will not be modelled by a collapsing thin shell  of pressureless null dust. Rather, it is more appropriate to be modelled
by a collapsing thin shell of charged fluid with some non-zero pressure.
Here, we will compute the two-point correlation function and the expectation value
of the Wilson loop operator on the boundary field theory by probing the bulk with the geodesic and minimal area surfaces 
respectively. We find that the thermalization is always top-down and for a fixed value of the characteristic probe length $l$,
the thermalization time decreases as one increases the value of $\gamma$, i.e., the QGP with a higher value of the Weyl coupling 
would thermalize faster. Further, we elaborate upon several interesting properties of the theory with a Weyl correction, for example the
appearance of a swallow-tail pattern in the thermalization curve that can be controlled by $\gamma$. 

The paper is organized as follows: In section 2 we construct the black brane solution in linear order in the Weyl coupling 
constant. In section 3 we start our study of the holographic thermalization and set up the corresponding AdS-Vaidya solution by
modelling the dynamical gravity with a thin shell of charged matter. In section 4 we discuss in detail about the two non-local 
observables we would probe: the two point correlation function and the Wilson loop. In section 5 we give a detailed description
of the numerical procedure and explain the effect of the Weyl coupling constant and the chemical potential on the thermalization.
In section 6 we summarize our main results.  Three appendices at the end of the paper provide material supplementary to the 
main text. 

\section{Black Brane Solution with Weyl Corrections}

In this section, we first write down the model action and then construct the black brane solution solving the Einstein and Maxwell
equations. We consider the following action where a five-dimensional gravity with a negative cosmological constant is coupled 
to a $U(1)$ gauge field $A$ by the following two and four-derivative interactions :
\begin{eqnarray}
&&\textit{S} = \frac{1}{16 \pi G_5} \int \mathrm{d^5}x \sqrt{-g} \ \ \bigl[R+\frac{12}{L^{2}}
-\frac{1}{4}\textit{F}_{\mu\nu}\textit{F}^{\mu\nu}+ L^2 (c_1\textit{R}_{\mu \nu \rho \lambda}
\textit{F}^{\mu\nu}\textit{F}^{\rho\lambda} \nonumber\\
&&+c_2 R_{\mu\nu}\textit{F}^{\mu}\hspace{0.1mm} _{\rho}\textit{F}^{\nu\rho} + c_3 R F_{\mu\nu}F^{\mu\nu})\bigr] ,
\label{actiongeneral}
\end{eqnarray}
where $F=dA$ is the usual Faraday 2-form. The Maxwell term $F_{\mu\nu}F^{\mu\nu}$ represents the familiar two derivative
interaction whereas the coefficients $c_1$, $c_2$ and $c_3$ represent the coupling constants for the four derivative interaction
terms which couple two derivatives of the gauge field to the spacetime curvature. $L$ symbolizes the AdS length which is related to the 
cosmological constant $\Lambda$ by $\Lambda=-{6\over L^2}$. We will work in a unit where $16\pi G_5=1$, $G_5$ being the five dimensional
gravitational constant. The factor $L^2$ in the four-derivative interaction terms is brought in to make the coefficients  
$c_1$, $c_2$ and $c_3$ dimensionless.

The action above is phenomenological in nature and let us briefly elaborate on this. As pointed out in the introduction, the work 
of \cite{Myers} started with the most general four derivative action of gravity with a single $U(1)$ gauge field. It was
shown in that paper that by choosing proper field redefinition, the action can in fact be written in terms of a fewer number of terms. 
The final action in that paper contained five four-derivative interaction terms proportional to 
$R_{\mu\nu\rho\lambda}R^{\mu\nu\rho\lambda}$, $R_{\mu\nu\rho\lambda}F^{\mu\nu}F^{\rho\lambda}$, $(F^2)^2$, $F^4$ and 
$\epsilon^{\mu\nu\rho\lambda\sigma}A_{\mu}R_{\nu\rho\alpha\beta}R_{\lambda\sigma}\ ^{\alpha\beta}$ along with the standard 
two derivative terms for five dimensional gravity along with a Chern-Simons term, where, $F^2=F_{\mu\nu} F^{\mu\nu}$ 
and $F^4=F^{\mu}\ _{\nu}F^{\nu}\ _{\rho}F^{\rho}\ _{\lambda}F^{\lambda}\ _{\mu}$. In \cite{Anninos}, an alternative field redefinition
was used, which in turn retained the $R^2$ and $R F^2$ in the action. One can, in principle, understand thermalization with 
all five generic terms turned on in the action, but this will be complicated, especially since we expect the physics to depend strongly
on relative values of the coefficients. We will rather focus on one particular type of correction. Our approach here is to write an action involving the
coupling of the curvature to the gauge field. As pointed out earlier, such terms are sufficient to study charge transport properties
of the dual CFT. We treat this as a phenomenological model to study thermalization in the dual field theory. 

We consider a linear combination \cite{Ritz}, \cite{Sachdev}, \cite{Cai} of the three four-derivative interaction
terms of (\ref{actiongeneral}) to express it into a simple form :
\begin{equation}
\textit{S} = \frac{1}{16 \pi G_5}\int \mathrm{d^5}x \sqrt{-g} \ \ \bigl(R+\frac{12}{L^{2}}
-\frac{1}{4}\textit{F}_{\mu\nu}\textit{F}^{\mu\nu}+\gamma L^2 \textit{C}_{\mu \nu\rho\lambda}
\textit{F}^{\mu\nu}\textit{F}^{\rho\lambda}\bigr) ,
\label{action}
\end{equation}
where the five dimensional Weyl tensor $\textit{C}_{\mu \nu\rho\lambda}$ is given by 
\begin{eqnarray}
 \textit{C}_{\mu \nu\rho\lambda}=\textit{R}_{\mu \nu\rho\lambda}+{1\over 3}(g_{\mu \lambda}R_{\rho \nu}+g_{\nu \rho} 
 R_{\mu \lambda}-g_{\mu \rho}R_{\lambda \nu}-g_{\nu \lambda}R_{\rho \mu})+{1\over 12}(g_{\mu \rho}g_{\nu \lambda}-
 g_{\mu \lambda}g_{\rho \nu})R .
\label{Weyl tensor}
\end{eqnarray}
Here, $\gamma$ represents the effective coupling for these higher derivative interaction terms. We will refer $\gamma$ as 
the `Weyl coupling' throughout the text. We will make a couple of remarks on the Weyl coupling here. 
In a five dimensional bulk AdS spacetime, it was shown by \cite{Ritz} that $\gamma$ is bounded, namely
$-{L^2 \over 16} \leq \gamma \leq {L^2 \over 24}$. While the lower bound arises to avoid the the possibility of superluminal propagation in the CFT 
by metastable quasi-particles, the upper bound appears to avoid the creation of certain ghost-like modes near the horizon. Hence in the probe 
limit, both signs of $\gamma$ are feasible. Unfortunately, in the present analysis which includes backreaction, we cannot establish such a rigorous
bound, as we will be working in a perturbative approximation up to first order in $\gamma$ (as we elaborate upon shortly). However, in the spirit 
of \cite{Ritz}, we will consider both signs of $\gamma$ in our numerical analysis. 

Before deriving the equations of motion one should note that the term $\textit{C}_{\mu \nu\rho\lambda}\textit{F}^{\mu\nu}
\textit{F}^{\rho\lambda}$ can be written in the  following form \cite{Myers},
\begin{eqnarray}
 \textit{C}_{\mu \nu\rho\lambda}\textit{F}^{\mu\nu}\textit{F}^{\rho\lambda}=\textit{R}_{\mu \nu \rho \lambda}
\textit{F}^{\mu\nu}\textit{F}^{\rho\lambda}-{4\over 3}R_{\mu\nu}\textit{F}^{\mu}\hspace{0.1mm} _{\rho}\textit{F}^{\nu\rho}
+{1\over 6}R F_{\mu\nu}F^{\mu\nu} .
\label{WeylFF}
\end{eqnarray}
Using the above form and making use of the Palatini identities (see Appendix A) we can construct the Einstein equation, 
\begin{eqnarray}
R_{\mu\nu}-{1\over2}g_{\mu\nu}R-{6\over L^2}g_{\mu\nu}-T_{\mu\nu}=0 \,,
\label{EinsteinEOM}
\end{eqnarray}
where $T_{\mu\nu}$ represents the energy-momentum tensor and has the following expression,
\begin{eqnarray}
 && T_{\mu\nu}={1\over 2}\bigl(g^{\alpha\beta}F_{\mu\alpha}F_{\nu\beta}-{1\over4}g_{\mu\nu}F_{\alpha\beta}F^{\alpha\beta}\bigr)+
 {\gamma L^2 \over 2}\bigl[g_{\mu\nu}C_{\delta\sigma\rho\lambda}F^{\delta\sigma}F^{\rho\lambda}-6 g_{\delta(\mu}
 R_{\nu)\sigma\rho\lambda}F^{\delta\sigma}F^{\rho\lambda} \nonumber \\
 &&+4\nabla_{\delta}\nabla_{\rho}(F^{\rho}\hspace{0.1mm}_{(\mu}F_{\nu)}\hspace{0.1mm}^{\delta})+{4\over3}\nabla^{\sigma}
 \nabla_{\sigma}(F_{\mu}\hspace{0.1mm}^{\rho}F_{\nu\rho})+{4\over3}g_{\mu\nu}\nabla_{\sigma}\nabla_{\delta}(F^{\delta}
 \hspace{0.1mm}_{\rho}F^{\sigma\rho})-{8\over3}\nabla_{\delta}\nabla_{(\mu}(F_{\nu)\rho}F^{\delta\rho}) \nonumber \\
 && +{8\over3}R_{\delta\sigma}F^{\delta}\hspace{0.1mm}_{\mu}F^{\sigma}\hspace{0.1mm}_{\nu}+{16\over3}R_{\sigma(\mu}
 F_{\nu)\rho}F^{\sigma\rho}-{1\over3}R_{\mu\nu}F^{\delta\sigma}F_{\delta\sigma}-{1\over3}g_{\mu\nu}\nabla^{\rho}
 \nabla_{\rho}(F^{\delta\sigma}F_{\delta\sigma}) \nonumber \\
 && +{1\over3}\nabla_{(\mu}\nabla_{\nu)}(F^{\delta\sigma}F_{\delta\sigma})-{2\over3}R g^{\delta\sigma}F_{\delta\mu}
 F_{\sigma\nu}\bigr] .
\label{EnergyMomentumTensor}
\end{eqnarray}
On the other hand, it is straightforward to write down the Maxwell equation,
\begin{eqnarray}
 \nabla_{\mu}(F^{\mu\lambda}-4\gamma L^2 C^{\mu\nu\rho\lambda}F_{\nu\rho})=0 .
 \label{MaxwellEOM}
\end{eqnarray}
Now, taking into account the backreaction of the $U(1)$ gauge field on the spacetime, we wish to solve the above equations 
(\ref{EinsteinEOM}) and (\ref{MaxwellEOM}) and try to construct a planar black brane solution. We consider the following ansatz for
the metric \footnote{One can also start with a different metric ansatz following \cite{Myers} and construct the black brane 
metric (see Appendix B). But it turns out that this ansatz is not numerically efficient for our purposes of the present problem
and we would use the ansatz given by (\ref{metric ansatz}).} and the gauge field,

\begin{eqnarray}
ds^2&=&- \frac{r^2}{L^2} f(r) e^{-2\chi(r)}  dt^2+\frac{L^2}{r^2 f(r)}dr^2 +\frac{r^2}{L^2}(dx^2+dy^2+d\eta^2) \,,
\label{metric ansatz}
\end{eqnarray}
\begin{equation}
A = (\phi(r),0,0,0,0) \,.
\label{A ansatz}
\end{equation}

If we take into account the backreaction of the $U(1)$ gauge field, it is difficult to obtain an exact analytical expression
for the metric that is a solution to the Einstein and Maxwell equations. Hence, we will try to perturbatively solve those equations up to linear order
in $\gamma$. As we have mentioned, in a string theory, four derivative interactions are expected to arise as quantum corrections to a
two derivative action, and our assumption is thus reasonable. 
There is however a caveat here which we should elaborate upon. We found that obtaining a controlled perturbative expansion in $\gamma$ in the
context of holographic thermalization is difficult. In our numerical analysis, choosing appropriate small 
values of $\gamma$ might therefore seem a little arbitrary. Currently, we do not have a clear answer to this question. However, we have checked
that the essential qualitative features of our analysis remains unaltered if numerical values of $\gamma$ are chosen to be smaller than those considered 
in this paper. We thus expect our results to capture the essential physics up to first order in $\gamma$, with the chosen numerical values of the
Weyl coupling.  This is a drawback of our analysis and we will comment more on this towards the end of the paper. 

In order to obtain the black brane metric, we will follow the by now standard procedure in the literature. 
We consider the following forms for $f(r)$, $\chi(r)$ and $\phi(r)$
\begin{eqnarray}
f(r)&=&f_0(r)\bigl(1+\mathcal{F}(r)\bigr)\,,\nonumber\\
\chi(r) &=& \chi_0(r) + \chi_1(r)\,,\\
\phi(r) &=& \phi_0(r)+ \phi_1(r) .\ \nonumber
\label{perturbationansatz}
\end{eqnarray}

where $f_0(r)$, $\chi_0(r)$ and $\phi_0(r)$ are the zeroth order solutions representing a Reissner-Nordstr\"{o}m AdS black brane
and have the following exprssions,
\begin{eqnarray}
 f_{0}(r) &=& 1-\frac{M L^2}{r^4}+\frac{Q^2 L^2}{r^6} \,,\nonumber\\
 \chi_{0}(r) &=& 0 \,, \nonumber\\ 
  \phi_{0}(r) &=& {L^3\over 2} q \bigl({1\over r_h^2}-{1\over r^2}\bigr).
 \label{zeroth order soln}
\end{eqnarray}
Here, $M$ and $Q$ are integration constants related to the ADM mass and the charge of the black brane.
$q=\left(*F\right)_{xy\eta}=2\sqrt{3}{Q\over L^3}$ represents the charge density and $r_h$ denotes the position of the
event horizon of the black brane.

We deonte by $\mathcal{F}(r)$, $\chi_1(r)$ and $\phi_1(r)$, the $\mathcal{O}(\gamma)$ corrections which we get by solving 
(\ref{EinsteinEOM}) and (\ref{MaxwellEOM}) keeping the terms up to linear order in $\gamma$. These are given as
\begin{eqnarray}
 \mathcal{F}(r)&=& {\gamma \over f_0(r)} \bigl({r_h^4 \over r^4}k_1+{2 L^2 Q^2 k_2\over r^6}-{24 L^2 Q^2\over r^6}
 +{2 L^2 Q^2 \over r^6}k_4+{16 L^4 M Q^2 \over r^{10}}-{15 L^4 Q^4\over r^{12}}\bigr) \,,\nonumber\\
 \chi_1(r)&=&  \gamma \bigl(k_2-\frac{4 L^2 Q^2}{r^{6}}\bigr) \,,\\
 \phi_1(r)&=& \gamma \bigl(k_3-{\sqrt{3}Q\over r^2}k_4-{8\sqrt{3} L^2 M Q\over r^6}+{14\sqrt{3} L^2 Q^3 \over r^8}\bigr) .\ \nonumber
 \label{perturbations}
\end{eqnarray}
where $k_1$, $k_2$, $k_3$ and $k_4$ are dimensionless integration constants.
 
Our next task would be to determine these constants. Following \cite{Myers}, we evaluate them in a standard fashion by imposing
several constraints on the above equations. First, we note the asymptotic behaviour of the black brane metric,
\begin{eqnarray}
 ds^2 \vert_{r\rightarrow \infty}=- (f e^{-2\chi})_\infty dt^2+dx^2+dy^2+d\eta^2 .
    \label{CFT metric}
\end{eqnarray}
where, $(f e^{-2\chi})_\infty=\lim_{r\to\infty}f(r) e^{-2\chi(r)}$. This metric at $r\rightarrow \infty$ represents the background metric where 
the dual boundary CFT lives. Now, to fix the speed of light in the CFT to be unity, we then demand 
that $(f e^{-2\chi})_\infty=1$, which in turn gives $k_2=0$.

Now our second requirement is that the charge density $q$ remains unchanged. Note that, one can write the Maxwell equation 
(\ref{MaxwellEOM}) in the form $\nabla_{\mu}X^{\mu\lambda}=0$, where, $X^{\mu\lambda}$ is an antisymmetric tensor. Hence, its 
dual $(*X)_{x y\eta}$ is a constant and it is convenient to choose this constant to be the fixed charge density $q$, 
i.e., $(*X)_{x y\eta}=q$. Since the quantity $(*X)_{x y\eta}$ does not depend on $r$, we demand 
\begin{eqnarray}
 \lim_{r\rightarrow\infty}\left(*X\right)_{xy\eta}=q .
 \label{constraint1a}
\end{eqnarray}
On the other hand, we can always compute this quantity in the asymptotic limit as,
\begin{eqnarray}
 \lim_{r\rightarrow\infty}\left(*X\right)_{xy\eta}&=&\lim_{r\rightarrow\infty}\bigl[{r^3\over L^3} e^{\chi(r)}
 \left(F_{r t}-8\gamma L^2 C_{r t}\hspace{0.1mm}^{r t}F_{r t}\right)\bigr] \nonumber\\
 &=&\bigl(1 + \gamma k_4 \bigr)q .
\label{constraint1b}
\end{eqnarray}
Comparing (\ref{constraint1a}) and (\ref{constraint1b}), we obtain $k_4=0$.

The third constraint is that we want to fix the position of the event horizon at $r=r_h$ for simplicity. Hence, we need
$f_0(r) \mathcal{F}(r) |_{r=r_h}=0$ which sets, 
\begin{eqnarray}
 k_1={10 L^2 M\over r_h^4}-{L^4 M^2\over r_h^8}-9 .
 \label{constraint2}
\end{eqnarray}
The final requirement is that we need $A_t$ has to vanish at the horizon, in order to have a well defined one-form for the gauge field
$A$. This implies $\phi_1(r_h)=0$, which in turn fixes the constant $k_3$,
\begin{eqnarray}
 k_3=-\frac{14 \sqrt{3} L^2 Q^3}{r_h^8}+\frac{8 \sqrt{3} L^2 M Q}{r_h^6} .
 \label{constraint3}
\end{eqnarray}

Since we have determined all the integration constants we can write down the final expressions for $\mathcal{F}(r)$, $\chi_1(r)$ 
and $\phi_1(r)$ which we show here for completeness,

\begin{eqnarray}
 \mathcal{F}(r)&=& {\gamma \over f_0(r)} \bigl(-{9 r_h^4 \over r^4}-{L^4 M^2\over r_h^4 r^4}+{10 L^2 M\over r^4}
 -{24 L^2 Q^2 \over r^6}+{16 L^4 M Q^2\over r^{10}}-{15 L^4 Q^4\over r^{12}} \bigr) \,,\nonumber\\
  \chi_1(r)&=&  - \gamma \frac{4 L^2 Q^2}{r^{6}} \,,\nonumber\\
  \phi_1(r)&=& 2 \sqrt{3} L^2 \gamma \bigl[4 M Q\bigl({1\over r_h^6}-{1\over r^6}\bigr)+
  7 Q^3 \bigl({1\over r^8}-{1\over r_h^8}\bigr)\bigr] .\ \nonumber
   \label{Final perturbations}
\end{eqnarray}
 
The Hawking temperature of this black brane is given by 
\begin{eqnarray}
 T = {1\over \pi L^2}\biggl(1-{Q^2 L^2\over 2r_h^6}\biggr)
  \biggl(r_h-{12 L^2 Q^2 \over r_h^5}\gamma \biggr)e^{{4L^2Q^2\over r_h^6}\gamma} .
  \label{Hawking Temp}
\end{eqnarray}
According to the gauge/gravity duality, it represents the temperature of the boundary field theory. Note that, for $\gamma=0$, 
it reduces to the Hawking temperature of the RNAdS black brane.
\begin{figure}[t!]
\centering
\includegraphics[width=4in,height=3in]{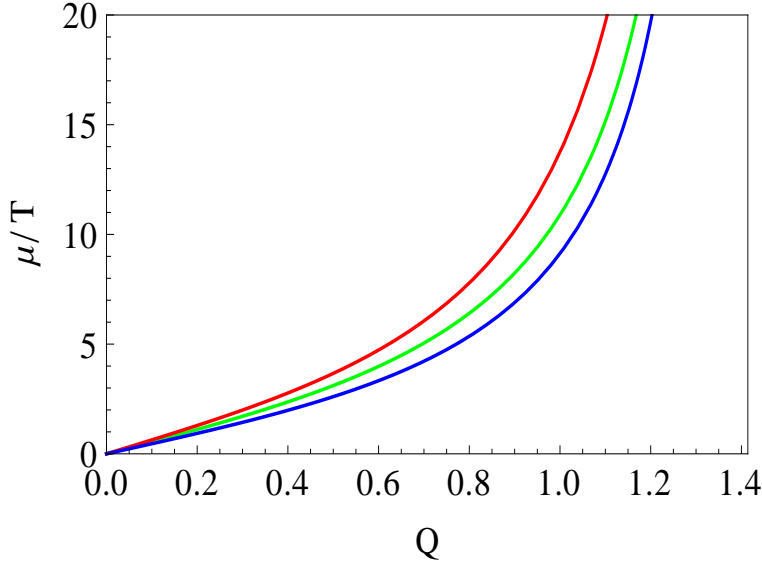}
\caption{\small ${\mu \over T}$ as a function of the black brane charge parameter $Q$ for different values of $\gamma$. The color
red, green and blue stands for $\gamma=0.02$, $0$ and $-0.02$ respectively. We have set $r_h=1$, $L=1$ and $\tilde{L}=1$.}
\label{mubyT}
\end{figure}
The condition $f(r_h)=0$ gives the relation between the charge $(Q)$ and the mass parameter $(M)$,
\begin{eqnarray}
 M={Q^2\over r_h^2}+{r_h^4\over L^2} .
\label{mass}
\end{eqnarray}
Demanding $T=0$, one can calculate the extremal charge of the black brane from (\ref{Hawking Temp}) as,
\begin{eqnarray}
 Q_{ext}=\sqrt{2}{r_h^3\over L} .
 \label{Qext}
\end{eqnarray}
The gauge/gravity duality suggests that the chemical potential $\mu$ of the boundary field theory should be identified with the
asymptotic value of the time component of the bulk gauge field, $\phi(r)$. On the boundary field theory $\mu$ has the dimension 
of energy (i.e., Length$^{-1}$). So, we redefine the gauge field as $\tilde{A_t}=A_t / \tilde{L}$ with some relevant scale 
`$\tilde{L}$' such that the chemical potential has the unit of energy, where $\tilde{L}$ has the dimension of length. Hence, 
the chemical potential on the boundary field theory is given by 
\begin{eqnarray}
 \mu = \lim_{r\rightarrow\infty}\tilde{A_t}=\lim_{r\rightarrow\infty}{\phi(r) \over \tilde{L}}={\sqrt{3}Q\over \tilde{L} r_h^2}
 +{2 \sqrt{3} L^2 Q\over r_h^6 \tilde{L}} \gamma \left(4 M-\frac{7 Q^2}{r_h^2}\right) .
 \label{chemical potential}
\end{eqnarray}
Now, in the boundary field theory we define a uesful dimensionless quantity,
\begin{eqnarray}
\frac{\mu}{T}=\frac{2\sqrt{3}\pi L^2 Q r_h^3}{\tilde{L}(2r_h^6-L^2 Q^2)}\biggl({6L^2Q^2\gamma-r_h^6-8\gamma r_h^6 \over 
12L^2Q^2\gamma-r_h^6}\biggr)e^{-{4L^2Q^2\over r_h^6}\gamma} .
 \label{mu by T}
\end{eqnarray}
We can explore the full range of ${\mu\over T}$, i.e., from ${\mu\over T}=0$ to 
${\mu\over T}=\infty$.
The case ${\mu\over T}=0$ corresponds to $Q=0$ and the other case ${\mu\over T}=\infty$ corresponds to $Q=Q_{ext}$ when the
Hawking temperature $T$ vanishes. Figure \ref{mubyT} shows the variation of ${\mu\over T}$ as a function of the black brane
charge parameter $Q$ for different values of the Weyl coupling constant $\gamma$. Here we have set the radius of the event
horizon $r_h=1$, the AdS length $L=1$ and the relevant scale $\tilde{L}=1$. The green curve corresponds to $\gamma=0$ when the Weyl corrected black brane 
reduces to a RNAdS black brane. The red and blue curves correspond to $\gamma=0.02$ and $-0.02$ respectively. For a small 
value of the charge parameter (say $Q<0.15$) the Weyl coupling $\gamma$ does not affect the ${\mu\over T}$ ratio. But for a 
sufficiently large value of the charge parameter the ${\mu\over T}$ ratio increases as $\gamma$ increases from a negative 
to a positive value. Since we are considering only the $\mathcal{O}(\gamma)$ correction to the metric and the gauge field, 
we will restrict ourselves to small values of $\gamma$ in all our numerical calculations. 

\section{Holographic Thermalization with Weyl Corrections}

In this section we briefly recapitulate the thermalization of the strongly coupled quantum field theory dual to the gravitational model introduced in the previous section. 
In particular, we will discuss how the new Weyl coupling term would affect such a thermalization process. We begin with a zero temperature state 
of a quantum field theory in four spacetime dimensions. According to the gauge/gravity duality, it is dual to a five dimensional pure AdS space.
Now, if we inject some energy to this zero-temperature state, it would evolve, and after a certain time reach the
thermal equilibrium at some non-zero finite temperature. This phenomenon is known as the thermalization. After the state 
thermalizes at some non-zero temperature, using the gauge/gravity duality, it would be identified with a charged black brane 
in AdS space. The Hawking temperature of this black brane represents the temperature of the final state in the dual field 
theory. So to describe the thermalization process in the field theory, we need to create an AdS black brane from a pure AdS 
space in the dual gravity sector. 

This formation of black branes is well-described in literature \cite{Bala}, \cite{Galante}, \cite{Kundu} by modeling the
spacetime with a AdS-Vaidya metric which desribes the collapse of a thin shell of charged matter from the boundary, into the 
bulk interior. The outer region of this collapsing shell represents a black brane metric while the inner one corresponds to 
a pure AdS spacetime. For our purpose, to study the effect of the Weyl coupling on thermalization, we need to construct a 
similar AdS-Vaidya type metric which at early time would correspond to a pure AdS space, and at late times would merge to the 
Weyl-corrected black brane metric after the shell collapses. Hence, we now focus on constructing the Weyl corrected AdS-Vaidya metric:

First, we introduce a new radial coordinate $z={L^2\over r}$. Note that the boundary is at $z=0$ and the horizon is 
at $z={L^2\over r_h}$ in terms of this new coordinate. Writing down the metric and the gauge field 
in the usual Eddington-Finkelstein coordinate 
\begin{eqnarray}
 dv=dt-{dz\over f(z) e^{-\chi(z)}} \,,
\end{eqnarray}
yields the following form,
\begin{equation}
 ds^2={L^2\over z^2}\biggl(-f(z) e^{-2\chi(z)} dv^2-2 e^{-\chi(z)} dv dz+dx^2+dy^2+d\eta^2\biggr)\,,
\label{metricEF}
\end{equation}
\begin{equation}
  A= \phi(z)\biggl(dv+{dz\over f(z) e^{-\chi(z)}}\biggr)\,.
\label{AEF}
\end{equation}
Although we have a $z$-component of the gauge field in the Eddington-Finkelstein coordinate, we can set $A_z=0$ through a gauge 
transformation. So, in this new coordinate system the gauge field becomes $A=A_t dv=\phi(z)dv$.

Now, instead of dealing with a constant mass $(M)$ and charge$(Q)$ parameter, if we assume them  to depend on the advanced time
coordinate $v$, i.e., $M=M(v)$ and $Q=Q(v)$, the functions  $f(z)$ and $\chi(z)$ in the metric would assume the form 
$f(v,z)$ and $\chi(v,z)$ respectively. Also, the gauge field $\phi(z)$ should be now written as $\phi(v,z)$. 
After considering $M$ and $Q$ as functions of the advanced time coordinate, the metric would no longer satisfy the Einstein and
Maxwell equation. We then need an external matter source to vary $M(v)$ and $Q(v)$, with the advanced time $v$. Considering this
external matter source, the Einstein and Maxwell equation can be written as,

\begin{eqnarray}
 R_{\mu\nu}-{1\over2}g_{\mu\nu}R-{6\over L^2}g_{\mu\nu}-T_{\mu\nu}= T_{\mu\nu}^{(ext)}\,, \nonumber \\
  \nabla_{\mu}(F^{\mu\lambda}-4\gamma L^2 C^{\mu\nu\rho\lambda}F_{\nu\rho})= J^{\lambda}_{(ext)} ,
 \label{EinsteinMaxwellVaidyaEOM}
\end{eqnarray}
where the external matter source must have the following expression for its non-zero components in order to obey the Einstein and Maxwell equation, 
 \begin{eqnarray}
 &&T_{v v}^{(ext)}=-\frac{3}{2} {z^3\over L^{10}} \bigl(2 z^2 Q(v) Q'(v)-L^4 M'(v)\bigr) -\frac{3}{2}{z^3\over L^{20}} \gamma
 \bigl[-2 {L^{16}\over r_h^4} M(v) M'(v) \nonumber \\
 &&+10 L^{14} M'(v)-32 L^4 z^6 M(v) Q(v) Q'(v)-16 L^{10} z^3 Q(v) Q''(v)+20 z^8 Q(v)^3 Q'(v) \nonumber \\
 &&-16 L^{10} z^3 Q'(v)^2+48 L^{10} z^2 Q(v) Q'(v)\bigr] ,\nonumber \\
 &&T_{v z}^{(ext)}= -72 \gamma  {z^5\over L^{10}} Q(v) Q'(v) =T_{z v}^{(ext)}, \nonumber \\ 
 &&T_{x x}^{(ext)}=T_{y y}^{(ext)}=T_{\eta\eta}^{(ext)}=-96 \gamma  {z^5\over L^{10}} Q(v) Q'(v). 
\label{EinsteinVaidyaEOM}
\end{eqnarray}
and
\begin{eqnarray}
J^{\lambda}_{(ext)}=-2 \sqrt{3} {z^5\over L^8} Q'(v) \bigl(1-4 \gamma  {z^6\over L^{10}} Q(v)^2\bigr)~\delta^{\lambda}_z.  
\label{MaxwellVaidyaEOM}
\end{eqnarray}
Here, the prime denotes a derivative with respect to $v$. Equation (\ref{MaxwellVaidyaEOM}) implies that 
$J^{\lambda}_{(ext)}J_{\lambda (ext)}=0$. Hence, the matter current, which sources the gauge field, is null which is permissible 
for a physical matter current (for related discussions, see \cite{Silva}). 

The stress-energy tensor $T_{\mu\nu}^{(ext)}$ can be diagonalized to determine the energy density and pressure. Solving the eigenvalue problem, 
$T_{\mu}^{\nu (ext)}n_{\nu}=\lambda \ n_{\mu}$ we get the energy density, $\rho=72 \gamma  \frac{z^7 q(v) q'(v)e^{\chi(v,z)}}{L^{12}}$ with 
$n^{\mu}=(0, -{z^2\over L^2} e^{\chi(v,z)},0,0,0)$ as the corresponding eigenvector, which is null.  This implies that, the stress energy tensor has 
support along the lightlike direction. In fact, the stress energy tensor can be written using two linearly independent null vectors
$n_1^{\mu}=(0, -{z^2\over L^2} e^{\chi(v,z)}, 0, 0, 0)$ and $n_2^{\mu}=(-1, {1\over 2} f(v,z) e^{-\chi(v,z)}, 0, 0, 0)$, with $n_1.n_2=-1$. Hence, the characteristic
surfaces of the Weyl-corrected Einstein and Maxwell equations are lightcones, which supports our analysis considering lightlike collapse.

However, (\ref{EinsteinVaidyaEOM}) reveals that the infalling matter we are dealing
with does not represent a shell of pressure-less null dust as considered in the works of \cite{Bala} - \cite{Abdalla}. 
Rather, we now have a thin shell of charged null fluid having finite pressure.\footnote{These types of solutions
have been well studied, see e.g \cite{Husain}.} Therefore, we will consider
the collapse of this null shell of fluid into a black brane to study the thermalization. 
Here, we should mention that the authors of \cite{Baron} considered the evolution of thin shells made of different kinds of degrees of freedom,
to study the dynamics of thermalization. These degrees of freedom are governed by different equations of state. It was shown there that the shells move and 
collapse with different velocities depending on the equation of state. In a similar manner one can in principle start with our Weyl-corrected Vaidya geometry, 
and taking a particular equation of state (e.g., $p=a \rho$ with $p$ and $\rho$ being the pressure and energy density, respectively, with in the shell and $a$ is a constant),
one can, compute the shell velocity and study the collapse using the Israel junction condition \cite{Israel}, \cite{Poisson}. However, in this paper,
we are interested in the effect of the Weyl-coupling constant, rather than different equations of state, or the degrees of freedom that constitute the shell. 

Now, we must make sure that the 
energy-momentum tensor supporting the time dependent bulk solution satisfies appropriate energy conditions, namely
the null energy condition (NEC): $T_{\mu\nu}^{(ext)}n_i^{\mu}n_i^{\nu}\geq 0$ $(i=1,2)$, where $n_1^{\mu}$ and $n_2^{\mu}$ are the null vectors 
that we have discussed above.\footnote{We mention here 
that \cite{kundu} has a nice description on why the NEC plays an important role in this context.} It can be checked that the NEC with $n_1^{\mu}$ is
satisfied trivially. On the other hand, replacing $n_2^{\mu}$ into the NEC, we have the following constraint
condition on the functions $f(v,z)$, $\chi(v,z)$ and $\phi(v,z)$,
\begin{eqnarray}
 T_{vv}^{(ext)}-f(v,z) e^{-\chi(v,z)} T_{vz}^{(ext)}\geq 0
 \label{NEC}
\end{eqnarray}
where, $T_{vv}^{(ext)}$ and $T_{vz}^{(ext)}$ are defined in (\ref{EinsteinVaidyaEOM}).

Now, we would choose the mass and the charge function in such a way that it can describe the thermalization process 
holographically, i.e., at early time $(v\rightarrow -\infty)$, $M(v)$ and $Q(v)$ would represent a pure AdS geometry while at 
late time $(v\rightarrow \infty)$ they would imply a black brane geometry. In particular, we would take two smooth functions 
which are often used in literature \cite{Bala}, \cite{Galante}, \cite{Kundu} for a numerical study of thermalization: 
\begin{eqnarray}
 M(v)&=&{M\over2}(1+\tanh{v\over v_0})\,,\nonumber \\
 Q(v)&=&{Q\over2}(1+\tanh{v\over v_0}) ,
 \label{mq}
\end{eqnarray}
where $v_0$ is the thickness of the null shell. Notice that, as $v\rightarrow -\infty$, $M(v)=Q(v)=0$ and the spacetime represents
a pure AdS geometry while as $v \rightarrow \infty$, $M(v)=M$, $Q(v)=Q$ and the spacetime reduces to a finite temperature 
black brane given by (\ref{metricEF}).

An useful physical situation would be the collapse of a shell having zero thickness $(v_0\rightarrow 0)$. Then 
one can assume that $M(v)=M \ \theta(v)$ and $Q(v)=Q \ \theta(v)$, where $\theta(v)$ is a step function. $v=0$ would represent
the position of the shell : for $v<0$ the spacetime would be a pure AdS space whereas for $v>0$ it would represent a black brane 
geometry. Substituting these forms of $M(v)$ and $Q(v)$ in (\ref{NEC}) one can check that the NEC is valid for any value 
of $v$ in this case. But one should be very careful on the validation of the NEC while performing numerical computations with 
the functions given in (\ref{mq}) i.e., with a shell of finite thickness $v_0$. We should choose the parameters $Q$, $M$ 
and $\gamma$ in such a way that the NEC is satisfied and in order to satisfy the NEC one can always tune the values of $L$ and
$r_h$. However, since the most physical situation would be the case with zero shell thickness, following \cite{Galante},
we scan the full parameter range i.e., from $Q=0$ to $Q=Q_{ext}=\sqrt{2}{r_h^3\over L}$.

\section{Non-local Observables}
In this section, we will pick out a set of observables to probe the thermalization in the four dimensional boundary field theory. 
Using the holographic principle, we would relate the particular observables to extended geometric objects in the bulk, 
representing the gravity dual for those. We would use the gravity duals to compute the particular observables and 
understand the thermalization. However, we cannot choose any local observable (e.g., expectation value of the energy-momentum 
tensor) to probe the thermalization. We have to compute some non-local observables (e.g., correlation function of the local 
operators) to understand the details of the thermalization process. In particular, we would like to calculate the two-point correlation 
function of the local gauge invariant operators at a fixed time and the expectation value of a rectangular Wilson loop  on the 
boundary field theory. The gauge/gravity duality provides a nice way to evaluate these non-local observables. The two-point 
correlation function can be interpreted as a renormalized geodesic length connecting the two points on the boundary. On the 
other hand, the expectation value of the rectangular Wilson loop corresponds to a minimal area surface extended into the bulk.

\subsection{Two-point Correlation Function}
We want to compute the equal time correlation function of local gauge invariant scalar operators $\mathcal{O}(t,x)$ with conformal
dimension $\Delta$ and study its evolution with time. As is well known, using the AdS/CFT correspondence and employing the saddle-point 
approximation for large values of the conformal dimension (i.e., $\Delta \gg 1$) one can evaluate this as,
\begin{eqnarray}
< {\cal{O}} (t,\textbf{x}) {\cal{O}}(t, \textbf{x}')>  \ \ \approx  e^{-\Delta {\cal{L}}} \, ,
\end{eqnarray}
where $\cal{L}$ is the length of the geodesic connecting the two points $(t,\textbf{x})$ and $(t,\textbf{x}')$ on the boundary
of the AdS space. Hence, for a very large value of the conformal dimension, $\cal{L}$ is proportional to the logarithm of the 
two-point correlation function. 

We consider two points $P(t,-{l\over2},y,\eta)$ and $Q(t,{l\over2},y,\eta)$ separated by a distance $l$ on the boundary field
theory and connect them to construct a spacelike geodesic which would extend into the bulk. From symmetry, it is clear that, the 
shape of the geodesic would depend only on the boundary spatial coordinate $x$, but not on $y$ or $\eta$. So we would treat $x$ as
the geodesic parameter. Then the solution to the geodesic equation is given by two functions $v(x)$ and $z(x)$ with the
following boundary conditions :
\begin{eqnarray}
 v(\pm l/2)  =  t \, , \,\,\,\,\,\,\,\, z(\pm l/2) = z_0 \, .
\end{eqnarray}
where $t$ is the time at the end point of the interval on the boundary and $z_0$ is an UV cutoff in the theory. 
Now, using (\ref{metricEF}), we can write down the length of the curve connecting the two points on the boundary as,
\begin{eqnarray}
\mathcal{L}_{curve}  =  \int \sqrt{-ds^2} = \int_{-l/2}^{l/2} dx
\frac{\sqrt{1-2 e^{-\chi(v,z)}v'(x)z'(x) -f(v,z) e^{-2\chi(v,z)} v'(x)^2}}{z(x)} \, .
\label{geodesiclength}
\end{eqnarray}
Here, the prime denotes a derivative with respect to $x$. The two functions $v(x)$ and $z(x)$ would minimize the length 
$\mathcal{L}_{curve}$ and yield the geodesic. Hence, we can think of the integrand of the above equation as the Lagrangian and 
$\mathcal{L}_{curve}$ as the action in the sense of classical mechanics. Note that, the integrand is not an explicit function of $x$, so there 
will be a conserved quantity,
\begin{eqnarray}
 \mathcal{H}  =  \frac{1}{z(x) \sqrt{1-2 e^{-\chi(v,z)}v'(x)z'(x) -f(v,z) e^{-2\chi(v,z)} v'(x)^2} } \, ,
 \label{Hamiltonian}
\end{eqnarray}
where, in analogy to classical mechanics, $\mathcal{H}$ corresponds to the Hamiltonian. Also note that, $x=0$ is the turning
point of the geodesics in a sense that the geodesic is symmetric around $x=0$. Now, we impose the initial conditions at $x=0$  :
\begin{eqnarray}
v(0)=v_* ,\ \ \,   z(0)=z_* ,\ \ \,  v'(0)=0 ,\ \ \,  z'(0) = 0 .\ 
\label{initial condition1}
\end{eqnarray}
Then (\ref{Hamiltonian}) gets simplified, since
\begin{eqnarray}
 \sqrt{1-2 e^{-\chi(v,z)}v'(x)z'(x) -f(v,z) e^{-2\chi(v,z)} v'(x)^2}={z_*\over z(x)} \, .
\label{Conservation}
\end{eqnarray}
By employing the Euler-Lagrange equation, we can write down the equations for $v(x)$ and $z(x)$ using (\ref{geodesiclength}) 
and (\ref{Conservation}),
\begin{eqnarray}
&& f(v,z)v''(x)+v'(x)z'(x)\partial_z f(v,z)+{v'(x)^2\over 2} \partial_{v}f(v,z)-2v'(x)z'(x)f(v,z)\partial_{z}\chi(v,z)\nonumber \\
&&-f(v,z)v'(x)^2\partial_{v}\chi(v,z)+e^{\chi(v,z)}\bigl(z''(x)-z'(x)^2 \partial_{z}\chi(v,z)\bigr)=0 \ ,\nonumber \\
 && 2+e^{-2\chi(v,z)}\bigl(z(x)v'(x)^2\partial_{z}f(v,z)-2f(v,z)v'(x)^2-2f(v,z)z(x)v'(x)^2\partial_{z}\chi(v,z)\bigr)\nonumber \\
 &&-2e^{-\chi(v,z)}\bigl(z(x)v''(x)+2v'(x)z'(x)-z(x)v'(x)^2\partial_{v}\chi(v,z)\bigr)=0 \ .
\label{geodesicVeomZeom}
\end{eqnarray}
This is a pair of second order non-linear coupled differential equations and difficult to solve analytically. However,
we can solve these numerically for different pairs of $(v_*,z_*)$, where we are given the initial conditions
(\ref{initial condition1}). Finally, we use (\ref{Conservation}) to write the on-shell geodesic length in a simple form,
\begin{eqnarray}
 \mathcal{L}  = 2 \int_{0}^{l/2} dx \frac{z_*}{z(x)^2} \equiv \mathcal{L} (l, t) .
\end{eqnarray}
Note that, $\mathcal{L}$ is a function of the separation $l$ between the two points and the time $t$ at those points. The $l$
dependence is clear from the integral, where as the $t$ dependence appears because of the factor $z_*$. For a particular 
initial value $(v_*,z_*)$, we get a particular time from the condition $v(\pm{l\over2})=t$. Now, if we vary $z_*$ we would 
get a different value of $t$. 

Notice that, the geodesic length $\mathcal{L}(l,t)$ has a divergent piece due to the contribution of the AdS boundary ($z=0$).
So one needs an UV cut-off (denoted by $z_0$ before) to get rid of this divergent piece. We define the relevant finite part of the 
geodesic length by taking out the divergent part in pure AdS geometry as
\begin{eqnarray}
 \mathcal{L}_{ren}(l,t) = \mathcal{L}(l,t) - 2 \ln ({2\over z_0})= 2 \int_{0}^{l/2} dx \frac{z_*}{z(x)^2} - 2 \ln ({2\over z_0}) \ .
 \label{renormalized geodesic length}
\end{eqnarray}
where, `$\mathcal{L}_{ren}$' is called the `renormalized geodesic length' which would represent the renormalized two-point 
correlation function on the boundary field theory.

\subsection{Wilson Loop}
We will use the Wilson loop as our second tool to probe the thermalization. We will explicitly calculate the expectation
value of the Wilson loop operator in the boundary field theory and study its time evolution. The Wilson loop is a non-local
gauge invariant observable which is defined as,
\begin{eqnarray}
 W(C)={1\over N}Tr (\mathcal{P}e^{\oint_C A_{\mu} \,dx^{\mu}})
\end{eqnarray}
where, the notation $\mathcal{P}$ denotes that we have a path-ordered integral of the gauge field $A$ over a closed loop $C$ and $N$ stands for 
the rank of the gauge group. Although it is in general difficult to compute the expectation value of the Wilson loop operator in a 
quantum field theory, it turns out to be easy using the AdS/CFT duality. It is related to the partition function in a
string theory as
\begin{eqnarray}
 <W(C)> = \int \mathcal{D}\Sigma \, e^{-\mathcal{S}_{NG}(\Sigma)}
\end{eqnarray}
where, $\Sigma$ denotes the world sheet extending into the bulk and having the closed loop $C$ as its boundary, i.e., 
$\partial \Sigma = C$. $S_{NG}(\Sigma)$ represents the string action, known as the Nambu-Gotto action. If the boundary field theory
is strongly coupled, we can make use of a saddle-point approximation and write the above partition function as
\begin{eqnarray}
 <W(C)>  \ \approx  e^{-{1\over 2\pi\alpha'}\mathcal{A}(\Sigma_0)}
\end{eqnarray}
where, $\alpha'$ denotes the inverse string tension and $\mathcal{A}(\Sigma_0)$ represents the area of the minimal surface 
world-sheet $(\Sigma_0)$ having the same boundary $C$.

Now we construct a spacelike rectangular Wilson loop $C$ on the $x-y$ plane of the AdS boundary having sides $l$ along the 
$x$-axis and $R$ along the $y$-axis and center at the origin. Further, if we assume translational invariance along the
$y$-axis, the shape of the bulk surface would depend only on $x$. Then the solution to the minimal area surface can again
be given by two functions $v(x)$ and $z(x)$ parameterized only by $x$. The same boundary conditions as in the earlier
case will be applicable here also,
\begin{eqnarray}
 v(\pm l/2)  =  t \, , \,\,\,\,\,\,\,\, z(\pm l/2) = z_0 \, .
 \label{bcWL}
\end{eqnarray}
Using (\ref{metricEF}) we can write down the area of the surface $\Sigma$ as,
\begin{eqnarray}
\mathcal{A}(\Sigma)  = R \int_{-l/2}^{l/2} dx
\frac{\sqrt{1-2 e^{-\chi(v,z)}v'(x)z'(x) -f(v,z) e^{-2\chi(v,z)} v'(x)^2}}{z(x)^2} \, .
\label{minimal area}
\end{eqnarray}
Like in the previous case of the geodesic, the integrand of the above equation has no explicit dependence on $x$. Hence, a
conserved quantity would be associated with it, namely,
\begin{eqnarray}
 \tilde{ \mathcal{H}_1}  =  \frac{1}{z(x)^2 \sqrt{1-2 e^{-\chi(v,z)}v'(x)z'(x) -f(v,z) e^{-2\chi(v,z)} v'(x)^2} } \, .
 \label{Hamiltonian 2}
\end{eqnarray}
Since the turning point of the minimal area surface is again at $x=0$ because of the symmetry of the problem, we can impose 
the initial conditions,
\begin{eqnarray}
v(0)=v_* ,\ \ \,   z(0)=z_* ,\ \ \,  v'(0)=0 ,\ \ \,  z'(0) = 0 .\ 
\label{initial condition2}
\end{eqnarray}
Using these initial conditions, (\ref{Hamiltonian 2}) simplifies to,
\begin{eqnarray}
 \sqrt{1-2 e^{-\chi(v,z)}v'(x)z'(x) -f(v,z) e^{-2\chi(v,z)} v'(x)^2}={z_*^2\over z(x)^2} \, .
\label{Conservation2}
\end{eqnarray}
We now minimize the area functional given by (\ref{minimal area}) and get the equations for $v(x)$ and $z(x)$. These are
similar to (\ref{geodesicVeomZeom}) but we show them here for completeness,
\begin{eqnarray}
&& f(v,z)v''(x)+v'(x)z'(x)\partial_z f(v,z)+{v'(x)^2\over 2} \partial_{v}f(v,z)-2v'(x)z'(x)f(v,z)\partial_{z}\chi(v,z)\nonumber \\
&&-f(v,z)v'(x)^2\partial_{v}\chi(v,z)+e^{\chi(v,z)}\bigl(z''(x)-z'(x)^2 \partial_{z}\chi(v,z)\bigr)=0 \ ,\nonumber \\
 && 4+e^{-2\chi(v,z)}\bigl(z(x)v'(x)^2\partial_{z}f(v,z)-4f(v,z)v'(x)^2-2f(v,z)z(x)v'(x)^2\partial_{z}\chi(v,z)\bigr)\nonumber \\
 &&-2e^{-\chi(v,z)}\bigl(z(x)v''(x)+4v'(x)z'(x)-z(x)v'(x)^2\partial_{v}\chi(v,z)\bigr)=0 \ .
 \label{WilsonVeomZeom}
\end{eqnarray}
We will solve these equations numerically the same way we do for the the two-point correlator. 
Now we use the conservation equation (\ref{Conservation2}) to express the area of the minimal surface as,
\begin{eqnarray}
\mathcal{A}(\Sigma_0)  =  2 R \int_{0}^{l/2} dx \frac{z_*^2}{z(x)^4} \equiv \mathcal{A} (l, t) \, .
\label{minimal area2}
\end{eqnarray}
Note that the minimal area surface is also divergent because of the $z=0$ contribution to the integral. So, we will work with 
a `renormalized' notion of the minimal area surface which we get by subtracting the divergent part. We define the `renormalized
minimal area surface' as,
\begin{eqnarray}
 \mathcal{A}_{ren}(l,t) = \mathcal{A}(l,t) - 2 {R\over z_0}=  2 R \int_{0}^{l/2} dx \frac{z_*^2}{z(x)^4} - 2 {R\over z_0} \ .
 \label{renormalized minimal area}
\end{eqnarray}
where, $z_0$ is the UV cut-off of the theory we are already familiar with.

\section{Numerical Results}
In this section, we provide the numerical details for computing the thermalization time in our Weyl corrected gravity model, 
and explain the results we get by probing the two-point correlation function and the Wilson loop on the boundary. 
For this purpose, it is necessary to separately solve the set of coupled differential equations (\ref{geodesicVeomZeom}) and (\ref{WilsonVeomZeom}). 
From now on, we would set the radius of the event horizon $r_h=1$, the AdS radius $L=1$, the UV cut-off $z_0=0.01$ and 
the shell thickness $v_0=0.01$ in all of our numerical calculations. 

For a particular value of the charge parameter $Q$, from (\ref{mass}), we have the mass parameter, $M=1+Q^2$. Now, for this 
fixed value of $M$ and $Q$, we first solve the pair of equations (\ref{geodesicVeomZeom}) subject to the initial conditions of
(\ref{initial condition1}). To extract the boundary time $t$, we fix the value of $v_*$ and tune the value of $z_*$ until we 
get $z=z_0=0.01$ at the end point of the geodesic. For example, setting $Q=0.5$ and $\gamma=0.02$, if we take the separation
of the two points on the boundary to be $l=3$ and fix the initial time $v_*=-1.42$, we get $z(\pm{l\over 2})=0.01$ 
just when the tip of the geodesic reaches the position $z_*=1.273155$. Thus we get the geodesic profiles $z(x)$ and $v(x)$ for
this particular value of $(v_*,z_*)$ and the corresponding boundary time would be computed as, $t=v(\pm {l\over 2})=0.0117278$. 
Fixing the boundary separation to be $l=3$, we can now take different values of $v_*$ and repeat this procedure to visualize
the geodesic profiles at different stages of time. Thus we obtain a number of geodesic profiles $v(x)$ and $z(x)$ at different
boundary times corresponding to different initial times $v_*$. 

We can calculate the renormalized geodesic length $\mathcal{L}_{ren}(l,t)$ from (\ref{renormalized geodesic length}) for 
all these geodesic profiles, and can study the behaviour of the renormalized geodesic length as a function of the boundary 
time. At this stage, it is convenient to introduce a dimensionless $l$-independent quantity ${\mathcal{L}_{ren}\over l}$
and plot  $\Delta \mathcal{L}=(\mathcal{L}_{ren}-\mathcal{L}_{thermal}) / l$ as a function of the boundary time $t$, where 
$\mathcal{L}_{thermal}$ is the thermal value of the renormalized geodesic length obtained by solving (\ref{geodesicVeomZeom}) 
with the final value of the mass and charge parameters, i.e., with $M(v)=M$ and $Q(v)=Q$.  

We use the same technique to solve the pair of equations (\ref{WilsonVeomZeom}) and view the profiles of $v(x)$ and $z(x)$ characterizing  
the minimal area surface for the rectangular Wilson loop. Like the previous case, we calculate the boundary time by using 
the same boundary conditions(\ref{bcWL}) and study the time evolution of the renormalized minimal area surface. As before, it is
convenient here to talk about a dimensionless renormalized quantity which does not depend on the area of the rectangular 
boundary Wilson loop. Hence, we will plot $\Delta \mathcal{A}=(\mathcal{A}_{ren}-\mathcal{A}_{thermal})/ R l$ as a function of 
the boundary time $t$, where $\mathcal{A}_{thermal}$ is the thermal value of the renormalized minimal area.

\begin{figure}
\centering
\subfigure[$\gamma=-0.01, t=0.527$]{
\includegraphics[width=0.3\columnwidth,height=0.2\columnwidth]{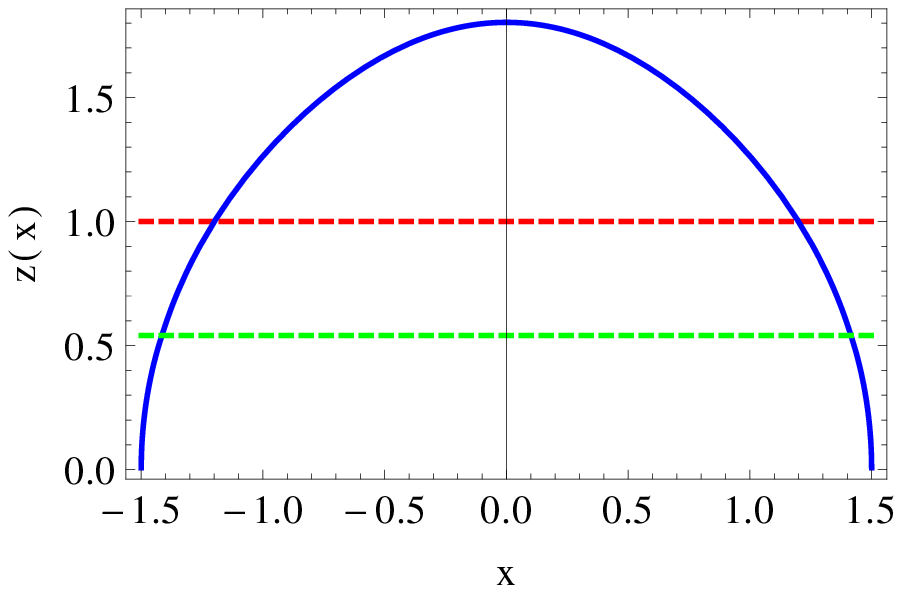}
\label{profile1} } 
\subfigure[$\gamma=0, t=0.527$]{
\includegraphics[width=0.3\columnwidth,height=0.2\columnwidth]{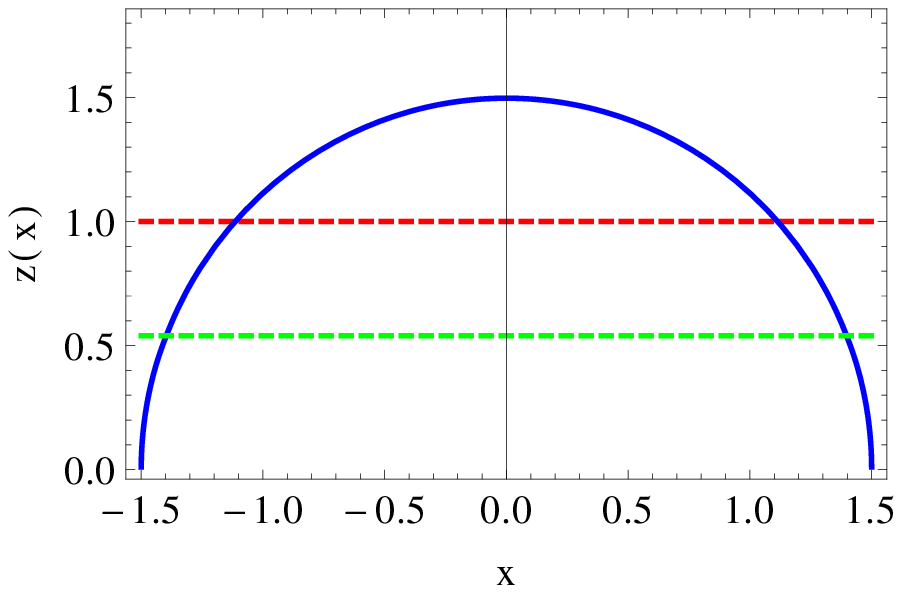}
\label{profile2} }
\subfigure[$\gamma=0.02, t=0.527$]{
\includegraphics[width=0.3\columnwidth,height=0.2\columnwidth]{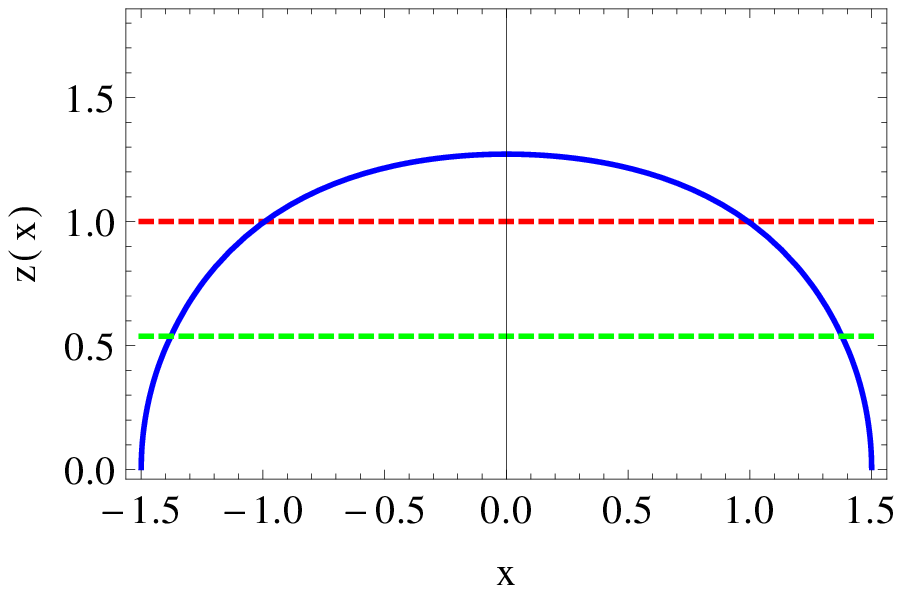}
\label{profile3} }
\subfigure[$\gamma=-0.01, t=0.800$]{
\includegraphics[width=0.3\columnwidth,height=0.2\columnwidth]{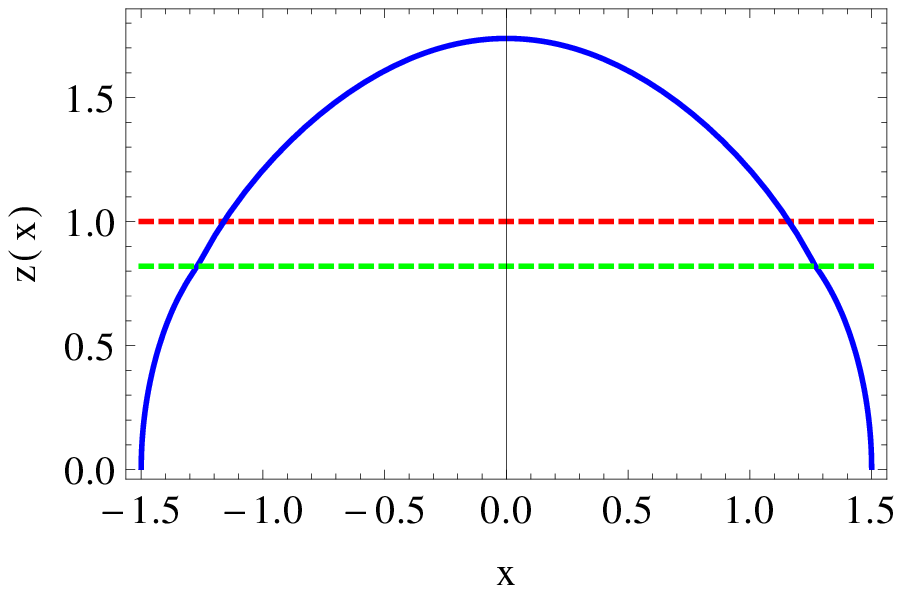}
\label{profile4} } 
\subfigure[$\gamma=0, t=0.802$]{
\includegraphics[width=0.3\columnwidth,height=0.2\columnwidth]{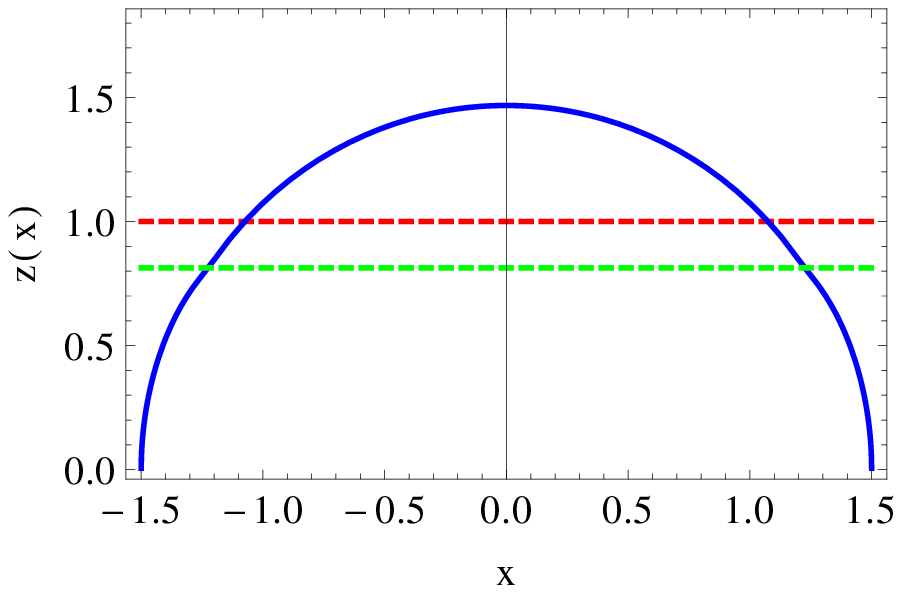}
\label{profile5} }
\subfigure[$\gamma=0.02, t=0.801$]{
\includegraphics[width=0.3\columnwidth,height=0.2\columnwidth]{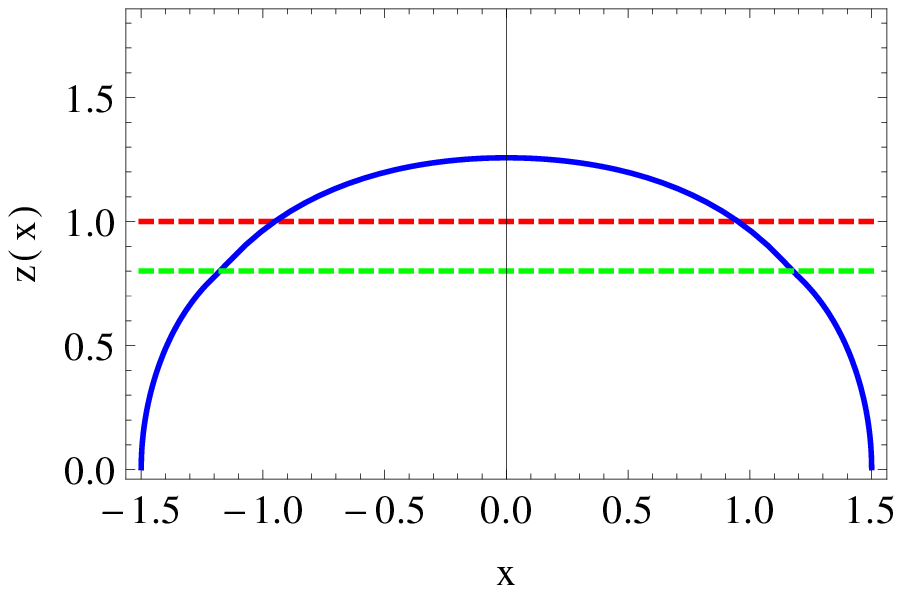}
\label{profile6} 
}
\subfigure[$\gamma=-0.01, t=1.001$]{
\includegraphics[width=0.3\columnwidth,height=0.2\columnwidth]{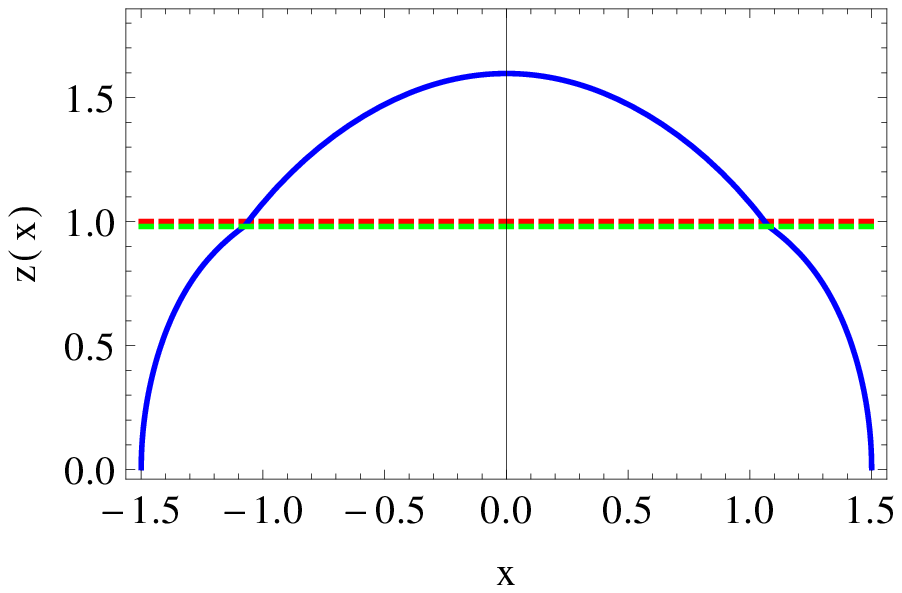}
\label{profile7} } 
\subfigure[$\gamma=0, t=1.003$]{
\includegraphics[width=0.3\columnwidth,height=0.2\columnwidth]{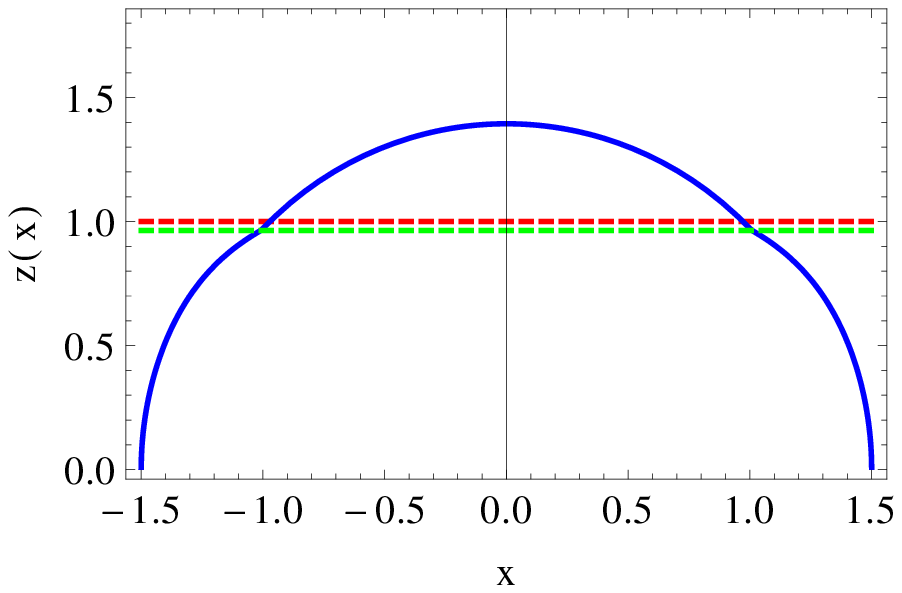}
\label{profile8} }
\subfigure[$\gamma=0.02, t=1.000$]{
\includegraphics[width=0.3\columnwidth,height=0.2\columnwidth]{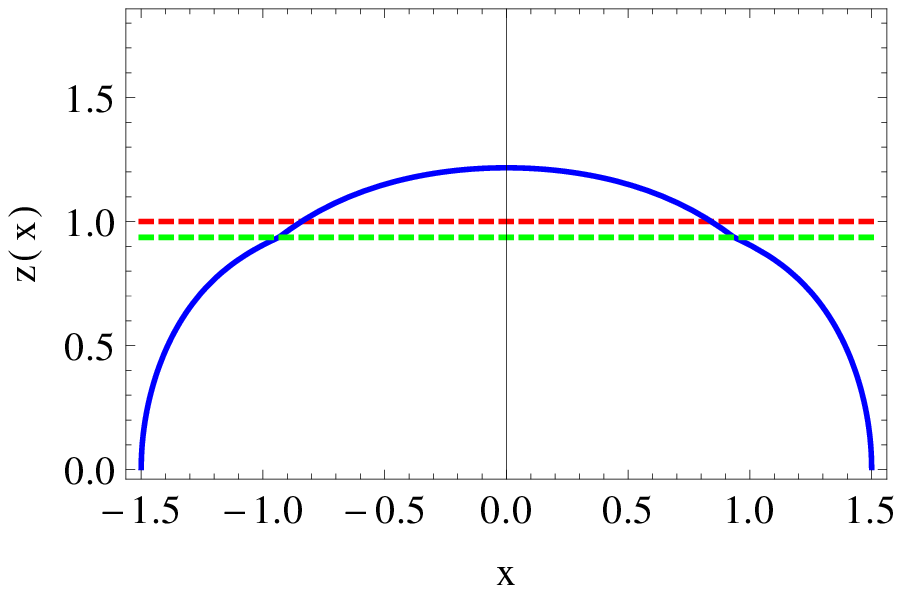}
\label{profile9} 
}
\subfigure[$\gamma=-0.01, t=1.348$]{
\includegraphics[width=0.3\columnwidth,height=0.2\columnwidth]{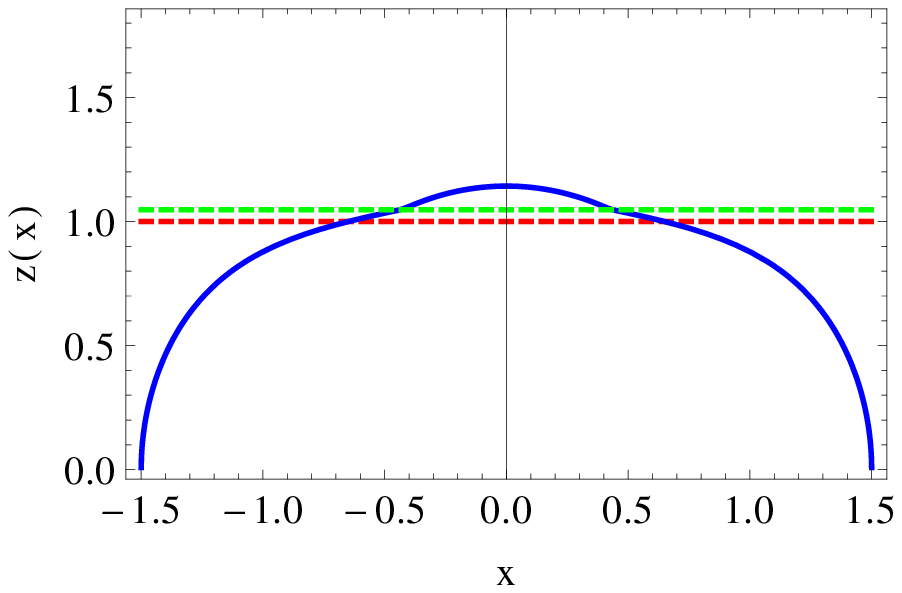}
\label{profile10} } 
\subfigure[$\gamma=0, t=1.354$]{
\includegraphics[width=0.3\columnwidth,height=0.2\columnwidth]{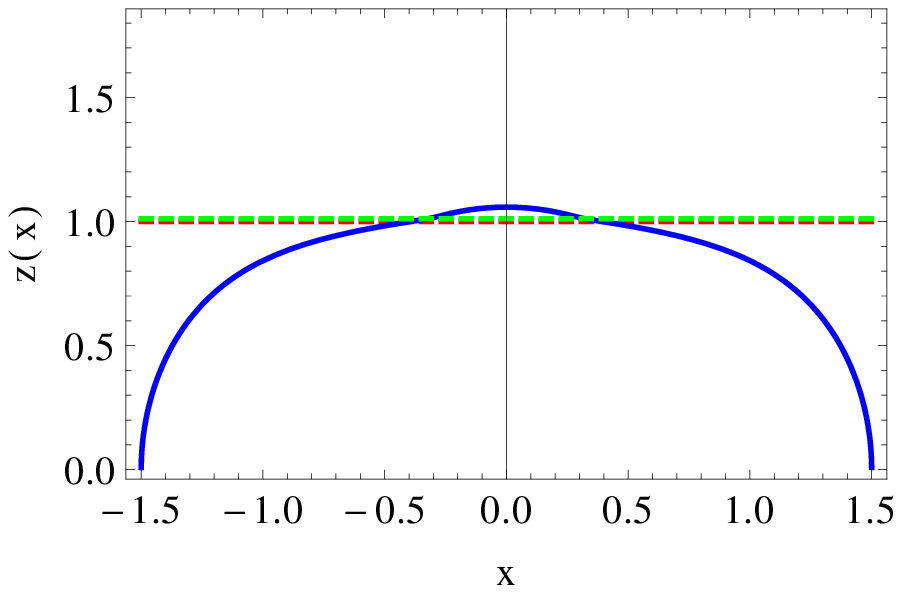}
\label{profile11} }
\subfigure[$\gamma=0.02, t=1.351$]{
\includegraphics[width=0.3\columnwidth,height=0.2\columnwidth]{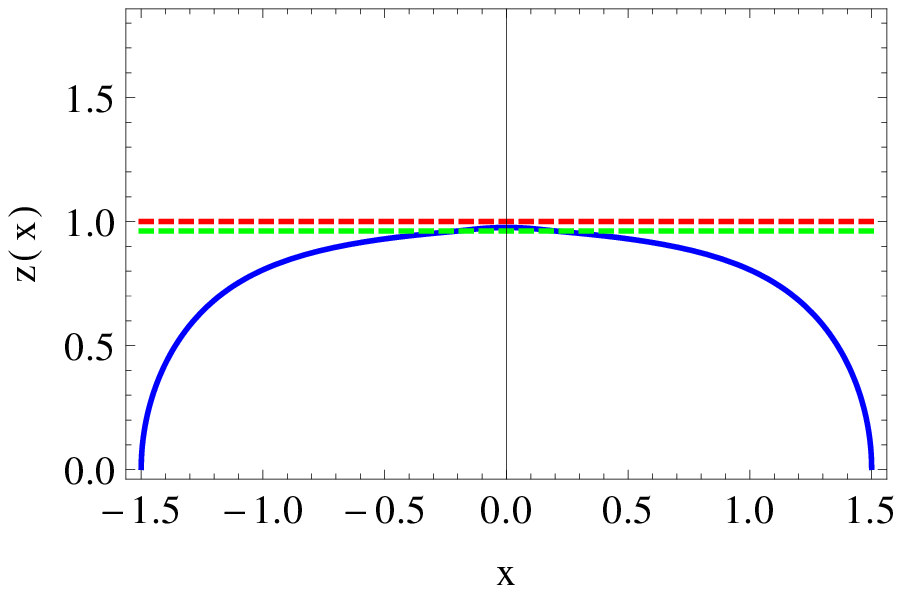}
\label{profile12} 
}
\caption{\small Time evolution of the geodesics and the position of the shell are shown for different values of $\gamma$ and 
fixed value of $Q=1$. The blue line denotes the geodesic profile at a particular boundary time, while the  green dashed line 
denotes the position of the shell at that particular time. The red dashed line at $z=1$ represents the horizon of the black 
brane to be formed at late time after the shell collapses. The separation between the boundary points is $l=3$ in all the 
cases.The left column corresponds to the time evolution of the geodesics for $\gamma=-0.01$, the middle one corresponds to
$\gamma=0$ and the right column represents the evolution for $\gamma=0.02$.}
\label{geodesic profile}
\end{figure}

\subsection{Two-point Correlation Function and the Renormalized Geodesic Length}

Here we study the time evolution of the two-point correlation function by probing the geodesic connecting those two points 
into the bulk AdS space. 
Figure \ref{geodesic profile} shows how the Weyl coupling parameter $\gamma$ affects the geodesic profiles at different 
stages of time and thus affects the thermalization. We have fixed the charge $Q=1$ and the separation between the boundary 
points $l=3$ and compared the geodesic profiles that appear at different times for different values of $\gamma$. The left column 
represents the time evolution of the geodesics for $\gamma=-0.01$, the middle one corresponds to $\gamma=0$ and the right 
column represents the time evolution for $\gamma=0.02$. Notice that as time elapses, the shell, shown by the dashed green line,
approaches $z=1$ (shown by the dashed red line) where the horizon of the black brane would be formed at late times. 
Since we have two different metrics on either side of the shell, whenever the geodesic crosses the shell, it gets refracted
by the shell. In the outer region of the shell, the geodesic propagates through an AdS black brane geometry whereas in the inner
region, it propagates through a pure AdS geometry. It is the refraction of the geodesic at the shell which makes the dual field
theory stray from thermality. As time elapses, the shell comes closer to the position where the event horizon would be formed
and the geodesic refraction at the shell becomes more prominent. The time when the geodesic no longer penetrates the shell 
defines the thermalization time since in this case the geodesic only propagates through the black brane geometry and hence 
on the dual field theory we have a thermal correlator.

Note that the geodesic with $\gamma=-0.01$ always penetrates `more' into the bulk than the other two geodesics. Now compare
the three figures of the last row at a fixed boundary time $t \approx 1.35$. With $\gamma=0.02$ the geodesic no longer crosses 
the shell. This suggests that the boundary system has thermalized at this value
of $\gamma$. But it still needs a small amount of time for $\gamma=0$ and an even larger amount of time for $\gamma=-0.01$ for 
the geodesic not to penetrate the shell. Hence, fixing the charge parameter $Q$, as we increase the Weyl coupling from a 
negative value to a positive value, we see that the boundary field theory takes less amount of time to thermalize. This is indicative 
of how a Weyl correction term can affect the thermalization time in the strongly coupled boundary field theory. 

\begin{figure}[t!]
\centering
\subfigure[$Q=0.5$]{
\includegraphics[scale=0.75]{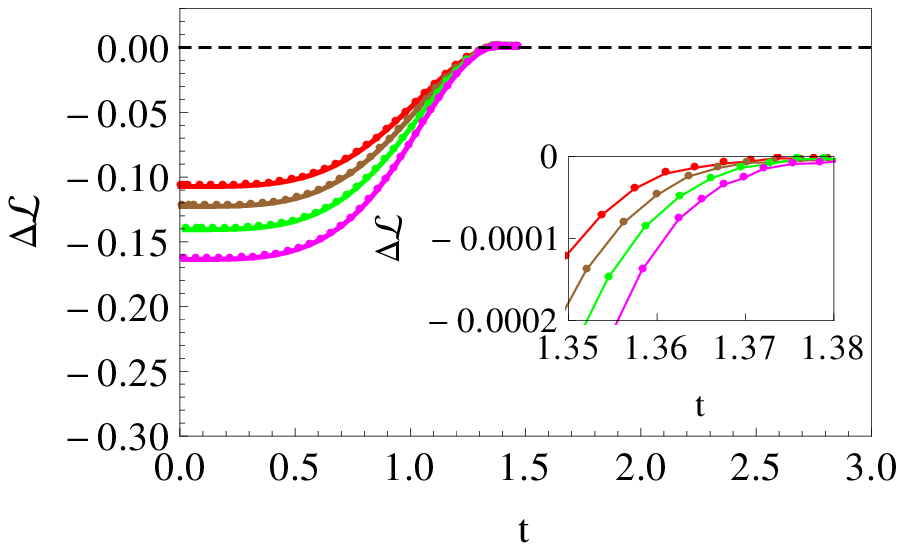}
\label{Geodesic Plot Q 0.5} } 
\subfigure[$Q=1$]{
\includegraphics[scale=0.75]{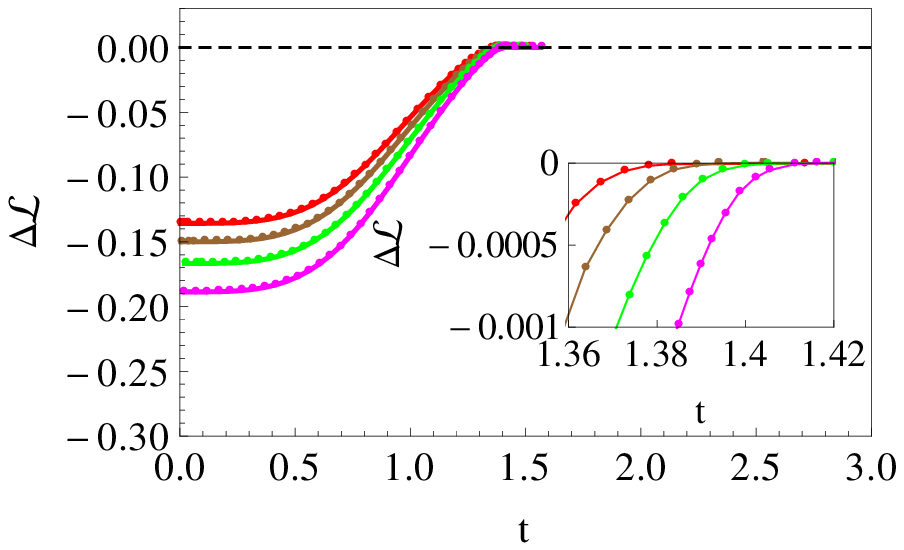}
\label{Geodesic Plot Q 1} }
\subfigure[$Q=\sqrt{2}$]{
\includegraphics[scale=0.75]{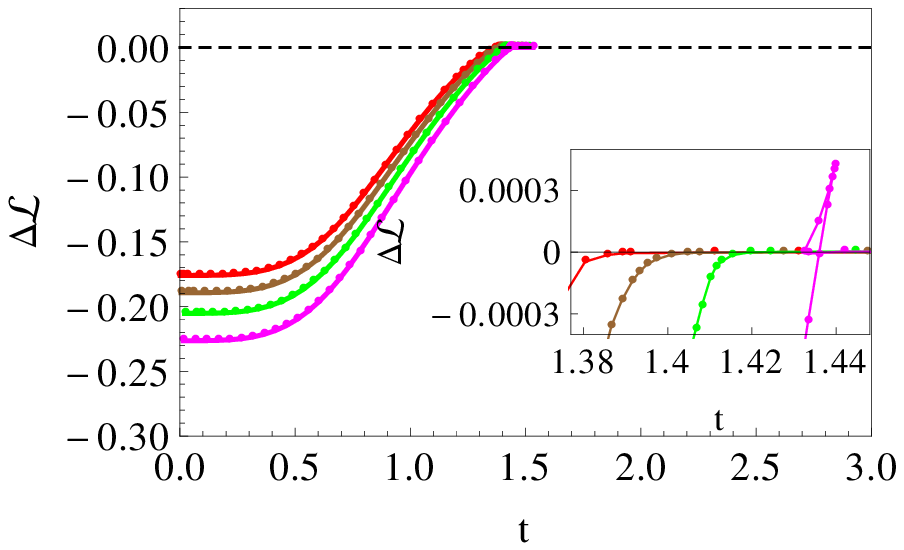}
\label{Geodesic Plot Q 1.414} }
\subfigure[$Q=\sqrt{2}$, $\gamma=-0.01$]{
\includegraphics[scale=0.75]{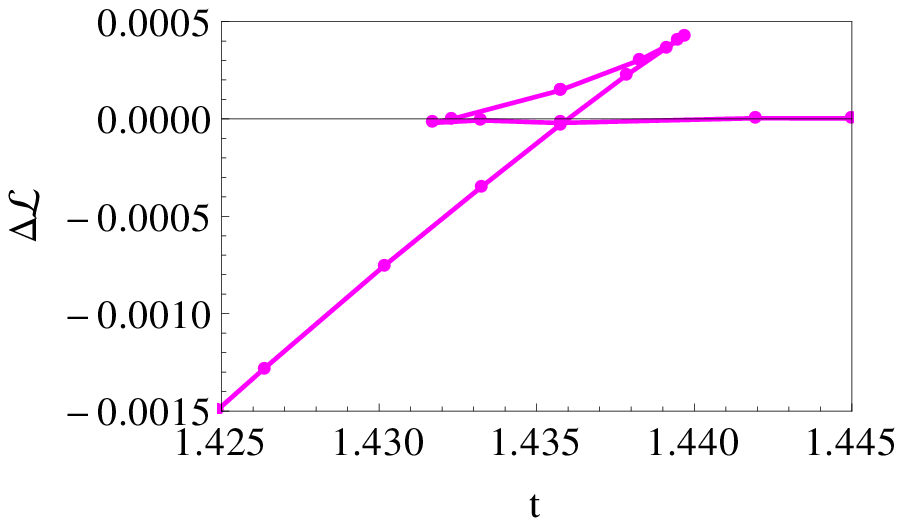}
\label{SwallowTail} }
\caption{\small Time evolution of the renormalized geodesic length for different values of $\gamma$ at fixed $Q$. (a) and (b)
correspond to the case $Q=0.5$ and $Q=1$ respectively while (c) represents the case with the extremal charge $Q=\sqrt{2}$. In
all cases we have fixed the separation between the two boundary points to be $l=3$. The curves with color
red, brown, green and magenta correspond to $\gamma=0.02, 0.01, 0$ and $-0.01$ respectively. (d) zooms the region where we get
the swallow-tail behavior of the thermalization curve with $Q=\sqrt{2}$ and $\gamma=-0.01$.}
\label{thermalization curves fixed Q probing geodesics}
\end{figure}

Now, we compute the dimensionless $l$-independent renormalized geodesic length $\Delta \mathcal{L}$ and plot it as a function of
the boundary time $t$ and generate the thermalization curves. We have already discussed the numerical method to calculate 
$\Delta \mathcal{L}(t)$ extensively and so here we directly show our results in figure \ref{thermalization curves fixed Q probing geodesics}.
This figure shows how the renormalized geodesic length and hence the two-point correlation function evolves with the boundary time $t$
with the Weyl coupling $\gamma$ as a parameter. 

In figures \ref{Geodesic Plot Q 0.5}, \ref{Geodesic Plot Q 1}, and 
\ref{Geodesic Plot Q 1.414} we have fixed $Q=0.5, 1$ and $\sqrt{2}$ respectively, while the separation between the two boundary points is
taken to be $l=3$. These figures show that, starting with a negative value, $\Delta \mathcal{L}$ increases with time and at a certain time
it saturates ending up at zero. But, at the beginning of the thermalization process, there is a delay, which was also reported by \cite{Bala}
and \cite{Galante}. They argued that the delay was because the boundary field theory experiences the sudden injection of energy only at a 
distance of the order of the thermal wavelength $\sim {1\over T}$. The time, when all the curves reach their corresponding thermal value, i.e.,
$\Delta \mathcal{L}(t)$ reaches zero, is referred to the thermalization time. Clearly, it sets a time scale for the Weyl corrected black brane
to form. We have zoomed the regions near the thermalization time to analyze the precise effect of $\gamma$ on the thermalization time.
 
These are shown in the insets of the corresponding figures. Notice that, when $Q$ is small, $\gamma$ has little effect on the thermalization 
time. But for a sufficiently large value of $Q$, as one tunes $\gamma$ from a negative value to a positive value, the thermalization time 
decreases which is also expected from the time evolution of the geodesics as shown in figure \ref{geodesic profile}. It is important to point 
that when $Q$ is very large, e.g., $Q=\sqrt{2}$, a swallow-tail pattern appears at the end of the thermalization curve with $\gamma=-0.01$,
whereas, with $\gamma=0.02$, $0.01$ and $0$ we have no such behaviour in the thermalization curve. One can also check that as $\gamma$ 
becomes more negative, the swallow-tail pattern becomes more prominent. This swallow-tail kind of behaviour has been reported
in \cite{Bala}, \cite{Galante}, \cite{Yang} and \cite{Johnson}  in the context of holographic thermalization, thermal and electromagnetic quenches.
In \cite{Bala} it was discussed that the emergence of the swallow-tail pattern depends on the dimension of the system while \cite{Galante} argued 
that it is rather a universal phenomenon which does not depend on the dimensionality of the AdS space. We should point out here that the 
swallow-tail behaviour appears for $\gamma=0$ for higher values of the charge $Q$, for large values of the boundary separation $l$. 
This seems to imply that this behaviour may not be related to potential causality violating issues as one switches on the Weyl coupling. 

\begin{figure}[h!]
 \centering
\subfigure[Time evolution of $z_{*}$]{
\includegraphics[scale=0.75]{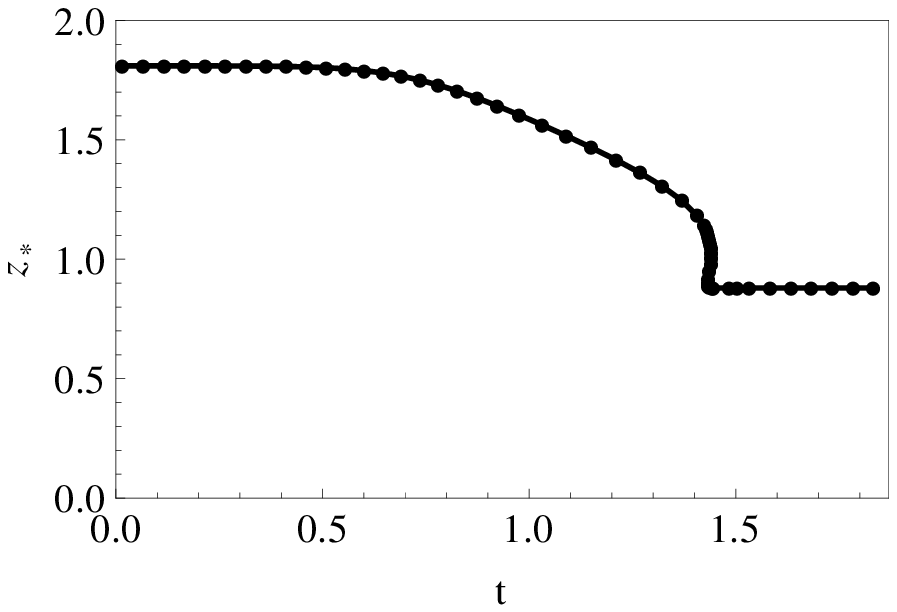}
\label{t0 vs zs} } 
\subfigure[Zoomed-in version of figure \ref{t0 vs zs}]{
\includegraphics[scale=0.75]{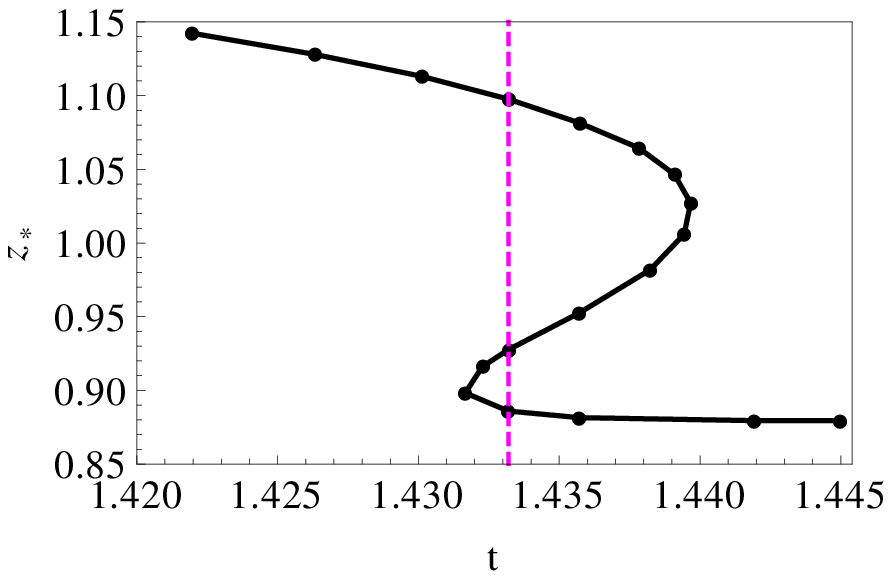}
\label{t0 vs zs zoomed} }
\label{Time evolution of zs}
\caption{\small (a) shows the time evolution of $z_*$ while (b) is the zoomed-in version of 
(a) at the time just before the thermalization. }
\end{figure}

In figure \ref{SwallowTail}, we present a zoomed view of the swallow-tail pattern. The pattern 
emerges because of the presence of three different geodesic profiles at a certain time before the thermalization and these 
three geodesics simultaneously extremize the bulk action at that particular time. Figure \ref{t0 vs zs} shows the time evolution of $z_{*}$,
which is zoomed and shown in figure \ref{t0 vs zs zoomed}. It clearly shows that between $t\approx 1.4315$ and
$t=1.44$, at any particular time, $z_*$ has three different values corresponding to the three different geodesics at that time.
Because of the multivaluedness of $z_*(t)$ one should be careful before applying the saddle-point approximation at 
the late time.

\begin{figure}[t!]
 \centering
 \subfigure[$v_*=-0.09032, z_*=1.098105$]{
 \includegraphics[scale=0.45]{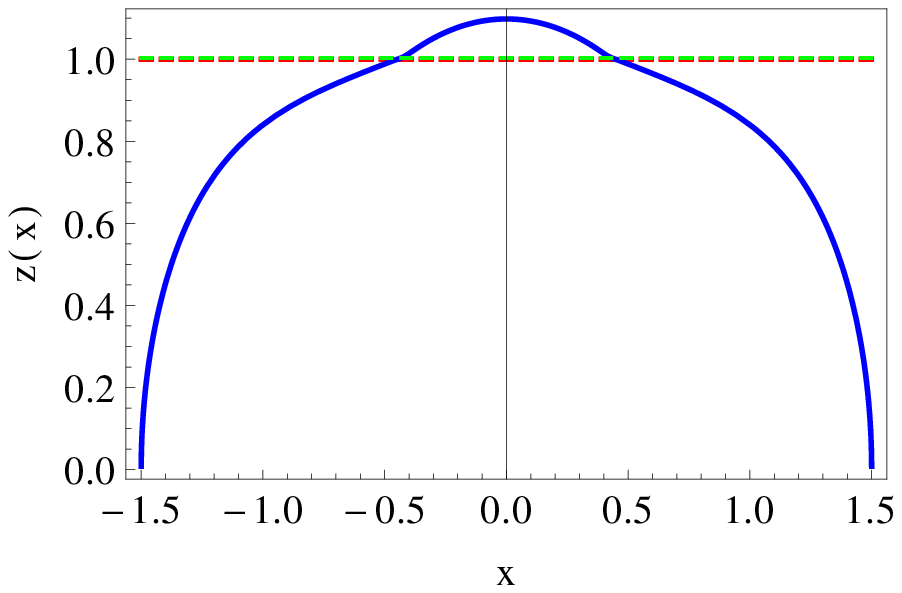}
 \label{magenta line top point} } 
 \subfigure[$v_*=-0.013, z_*=0.927644$]{
 \includegraphics[scale=0.45]{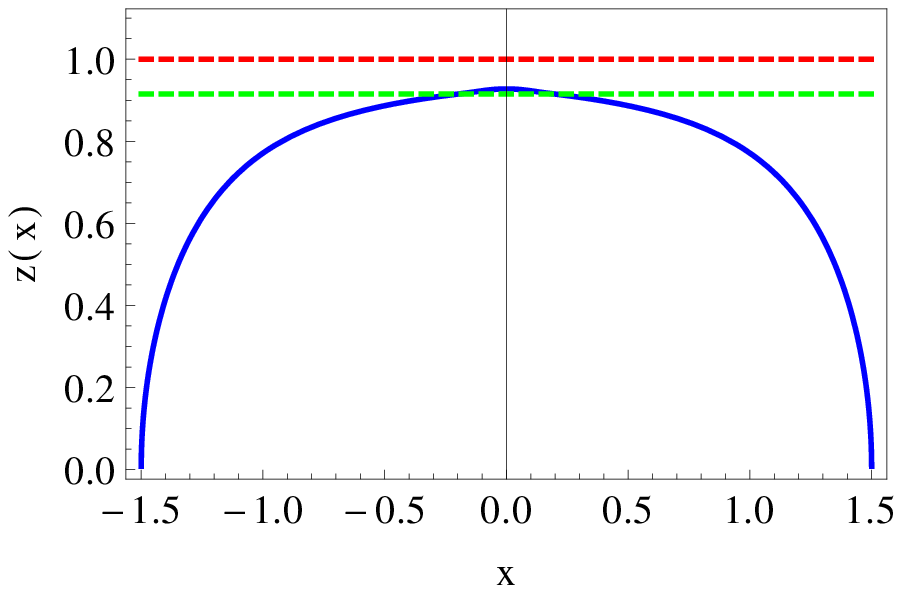}
 \label{magenta line middle point} }
 \subfigure[$v_*=-0.0001, z_*=0.885961$]{
 \includegraphics[scale=0.45]{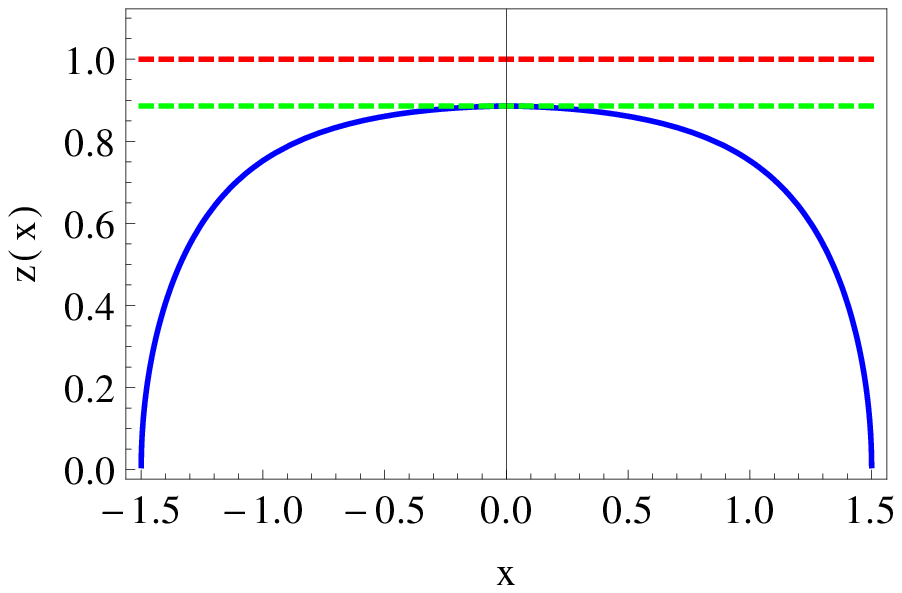}
 \label{magenta line bottom point} }
 \caption{\small The geodesic profiles from left to right correspond to the three points shown on the dashed magenta line in figure \ref{t0 vs zs zoomed}.
 The top point on the magenta line corresponds to the left figure, the middle point represents the middle figure
 and the bottom point on the magenta line corresponds to the right figure. The three geodesics extremize the action simultaneously
 at $t=1.4332$. We have fixed $Q=\sqrt{2}$, $\gamma=-0.01$ and the separation between the boundary points $l=3$.}
\end{figure}

Now consider the dashed magenta line in figure \ref{t0 vs zs zoomed} connecting the three points at $t=1.4332$. We have shown the three 
geodesics corresponding to these three points in figure 5, where the left figure corresponds to the top 
point on the magenta line in figure \ref{t0 vs zs zoomed}, the middle one corresponds to the middle point on the magenta line while
the right figure corresponds to the bottom point on the magenta line. Note that in figure \ref{magenta line top point} the geodesic 
crosses the shell and propagates inside the horizon while the shell is just inside the horizon. In figure \ref{magenta line middle point}
the shell is outside the horizon, the geodesic penetrates the shell but does not cross the horizon. In 
figure \ref{magenta line bottom point}, the shell lies well outside the horizon and in this case the geodesic does not cross the 
shell. 

In general, the appearance of a swallow-tail is usually an indicator of different scales present in the problem. In this case, if we follow the 
main curve, $\Delta {\mathcal L}$ has three branches before and beyond $t = 1.4357$ (the position of the kink in figure \ref{SwallowTail}). 
For $t=1.4332$, as we have just discussed, two of  the geodesics do not penetrate the horizon. This feature in fact continues beyond $t = 1.4357$,
up to $t \approx 1.439$, as is evident from figure \ref{t0 vs zs zoomed}. If we assume that the saddle point approximation remains valid in these regions, 
then this would seem to indicate two physical geodesic solutions corresponding to two different scales in the problem. From a field theory perspective
however, the two point correlator should be single valued, and this would suggest the breakdown of the saddle point approximation at late times. 
This issue is not resolved, and requires further investigation. 

\begin{figure}[t!]
 \centering
 \subfigure[$\gamma=-0.01$]{
  \includegraphics[scale=0.75]{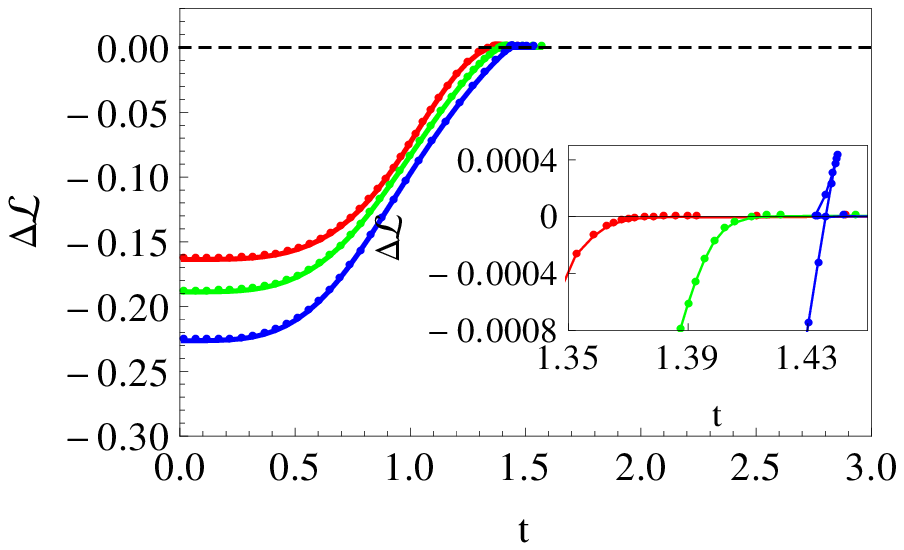}
  \label{Geodesic Plot Gamma -0.01} } 
  \subfigure[$\gamma=0.02$]{
  \includegraphics[scale=0.75]{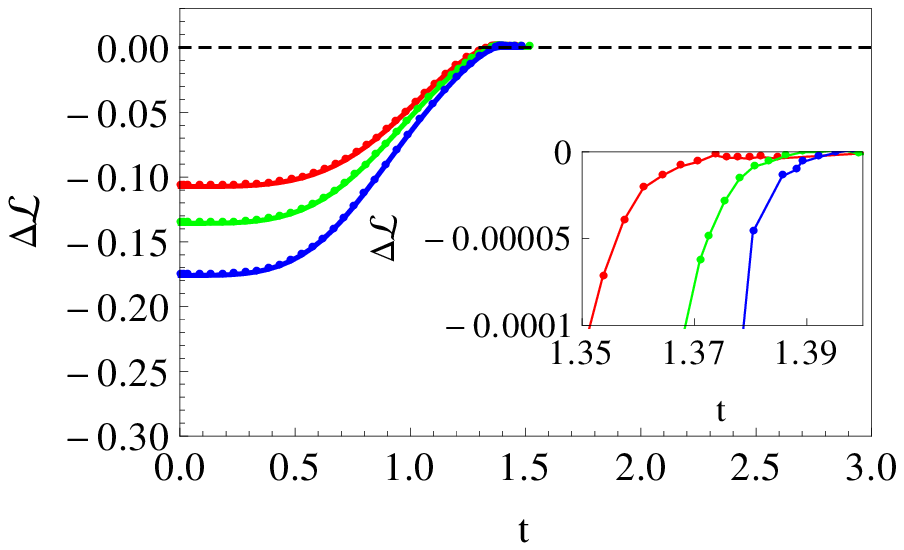}
  \label{Geodesic Plot Gamma 0.02} }
 \caption{\small Time evolution of the renormalized geodesic length for different values of $Q$ at fixed $\gamma$. The left one 
 corresponds to the case $\gamma=-0.01$ while the right one represents $\gamma=0.02$. The curves with color red, green and
 blue correspond to $Q=0.5, 1$ and $\sqrt{2}$ respectively.}
 \label{thermalization curves fixed Gamma probing geodesics}
\end{figure}

For completeness, we also provide the thermalization curves with $Q$ as a parameter at a fixed value of $\gamma$. These are 
shown in figure \ref{thermalization curves fixed Gamma probing geodesics}. From figure \ref{Geodesic Plot Gamma -0.01}, one can 
see that with $\gamma=-0.01$, the thermalization time increases as one increases the charge $Q$. In other words, if 
we change the ${\mu \over T}$ ratio from
zero to $\infty$, the thermalization time enhances. If we zoom the figure to view the behavior of the curves near the 
thermalization time, we notice that for the extremal charge $Q=\sqrt{2}$, which corresponds to ${\mu \over T}=\infty$, we 
get a swallow-tail kind of behaviour before the thermalization. For a larger value of $\gamma$ the change in the thermalization
time with $Q$ is negligible. Even in the inset of figure \ref{Geodesic Plot Gamma 0.02} we see a negligible change of the 
thermalization time within a very short region for $\gamma=0.02$. So, we would comment on it on after probing it by another
non-local observable, namely, the expectation value of the Wilson loop operator. But it is important to mention that we do not get 
any swallow-tail appearance in the thermalization curve for $\gamma=0.02$ even with the extremal charge $Q=\sqrt{2}$.

At this point, we define a time scale for the thermalization, $\tau_{crit}$, following \cite{Bala1}. It is the critical time 
when the peak of the geodesic touches the middle of the shell at $v=0$. This can be computed as,
\begin{equation}
 \tau_{crit}=\int_{z_0}^{z_*}{dz \over f(z)e^{-\chi(z)}}
 \label{critical time}
\end{equation}
where, $z_0$ is the UV cut-off and $z_*$ is the value of the $z$ coordinate at the peak of the geodesic. Note that, for a particular 
boundary separation $l$, $z_*$ would pick a particular value and substituting into the above formula we can determine $\tau_{crit}$
for that particular $l$.

\begin{figure}[h!]
 \centering
 \subfigure[$Q=1$]{
  \includegraphics[scale=0.75]{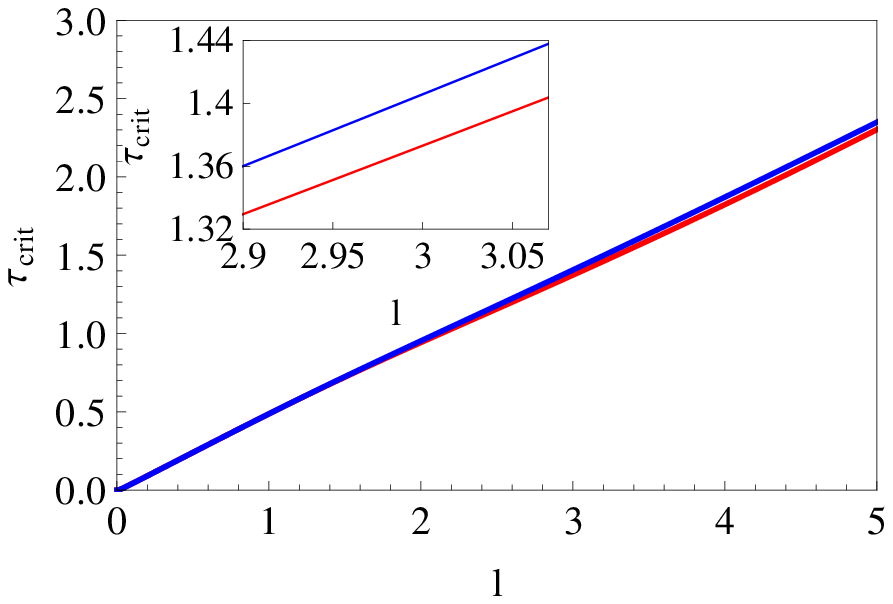}
  \label{Geodesic l vs tau Q1} } 
  \subfigure[$Q=\sqrt{2}$]{
  \includegraphics[scale=0.75]{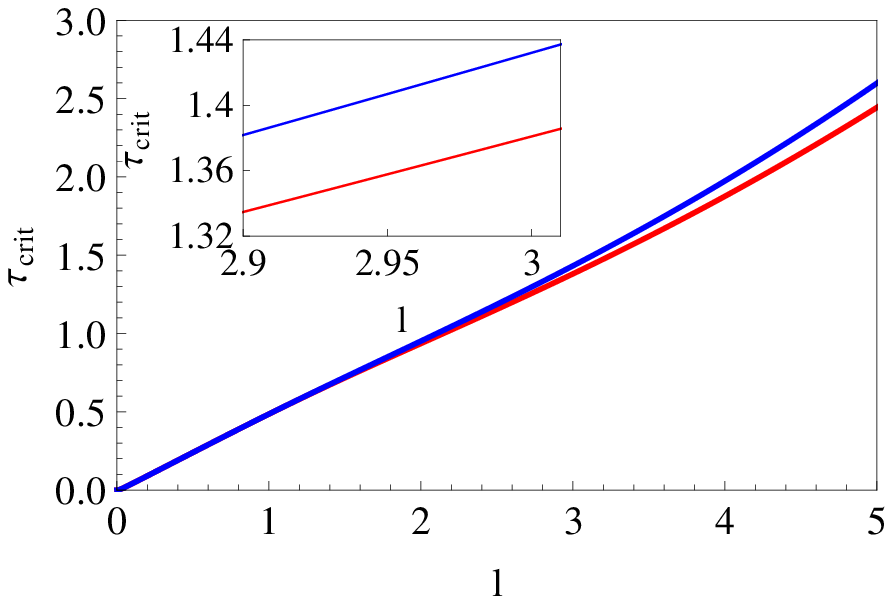}
  \label{Geodesic l vs tau Q1.414} }
 \caption{\small $\tau_{crit}$ as a function of $l$ for different values of $\gamma$ at fixed $Q$. The left one 
 corresponds to the case $Q=1$ while the right one represents $Q=\sqrt{2}$. The curves with color red and blue 
 correspond to $\gamma=0.02$ and $-0.01$ respectively.}
 \label{l vs tau curves fixed Q probing geodesics}
\end{figure}

In figure \ref{l vs tau curves fixed Q probing geodesics} the critical thermalization time  $\tau_{crit}$ is plotted as a function of
the boundary separation length $l$, which reveals that the thermalization is always top-down. A similar result was reported in 
\cite{Bala}, \cite{Galante}, \cite{Kundu}. This is not 
surprising and is a natural outcome of the dual geometrical probes we are using. If the boundary separation is small, the geodesic cannot
cross the shell and always propagate in the black brane geometry, yielding a thermal correlator on the boundary. But, if the boundary
separation is large enough, the geodesic would penetrate the shell and enter into the pure AdS geometry, thus the correlator would not be 
thermal and it would take a sufficient amount of time to be thermal. Hence $\tau_{crit}$ would be larger for a larger $l$.  
It is interesting to note that for small values of $l$,  $\tau_{crit}\sim {l\over 2}$, but as $l$ increases there is a deviation from
the linearity. From figure \ref{Geodesic l vs tau Q1.414} it is also clear that, for a sufficiently large value of the charge
parameter, the deviation occurs in $\tau >{l\over 2}$ for negative values of $\gamma$ and in $\tau <{l\over 2}$ for 
positive values of $\gamma$. The figure also shows that for very small values of $l$, $\tau_{crit}$ has a negligible dependence on 
$\gamma$, whereas, for a larger value of $l$, $\tau_{crit}$ decreases as one increases $\gamma$ at a fixed value of $Q$. Now fixing 
the value of $\gamma$, if one increases the charge parameter from $Q=1$ to $Q=\sqrt{2}$, $\tau_{crit}$ enhances. 

\begin{figure}
\centering
\subfigure[$\gamma=-0.01, t=0.527$]{
\includegraphics[width=0.3\columnwidth,height=0.2\columnwidth]{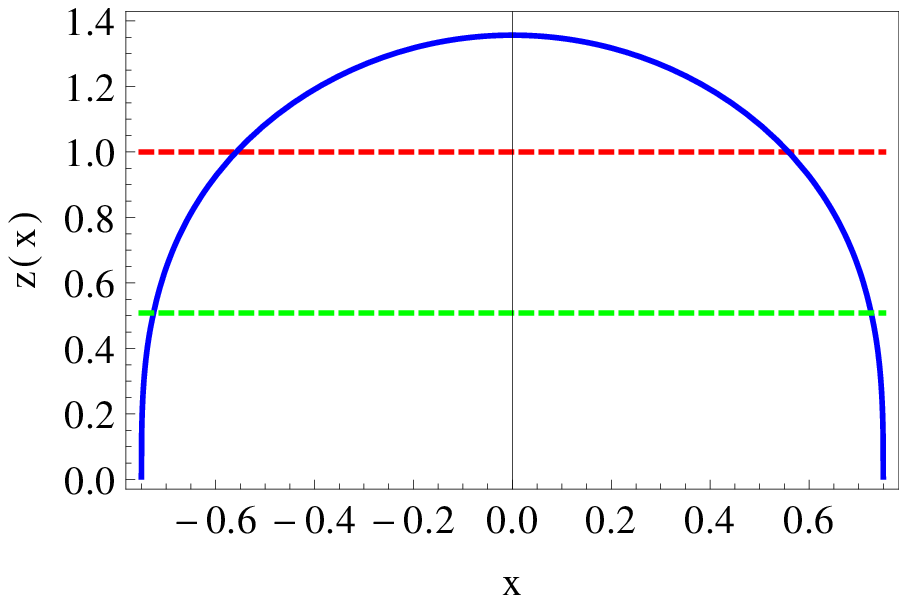}
\label{Wilsonprofile1} } 
\subfigure[$\gamma=0, t=0.527$]{
\includegraphics[width=0.3\columnwidth,height=0.2\columnwidth]{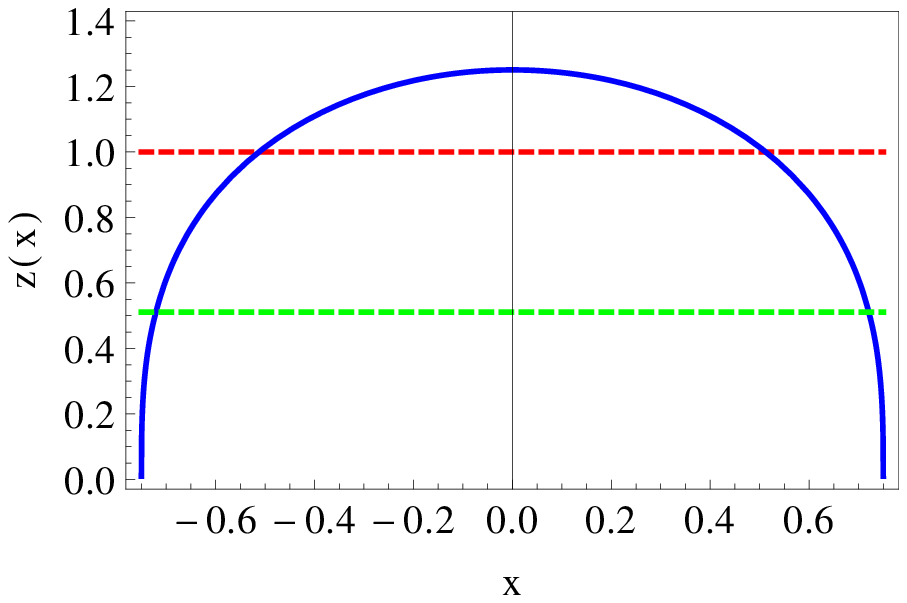}
\label{Wilsonprofile2} }
\subfigure[$\gamma=0.02, t=0.527$]{
\includegraphics[width=0.3\columnwidth,height=0.2\columnwidth]{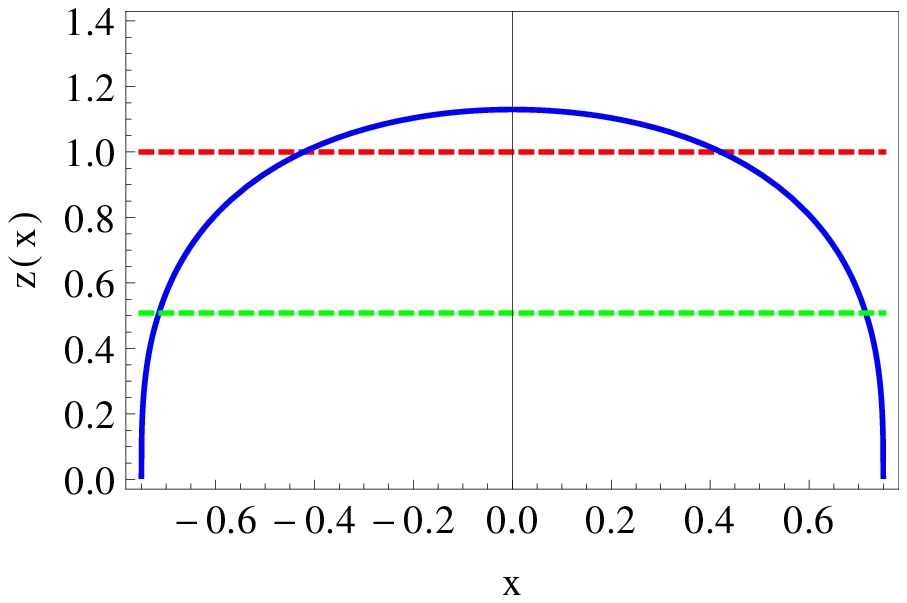}
\label{Wilsonprofile3} 
}
\subfigure[$\gamma=-0.01, t=0.800$]{
\includegraphics[width=0.3\columnwidth,height=0.2\columnwidth]{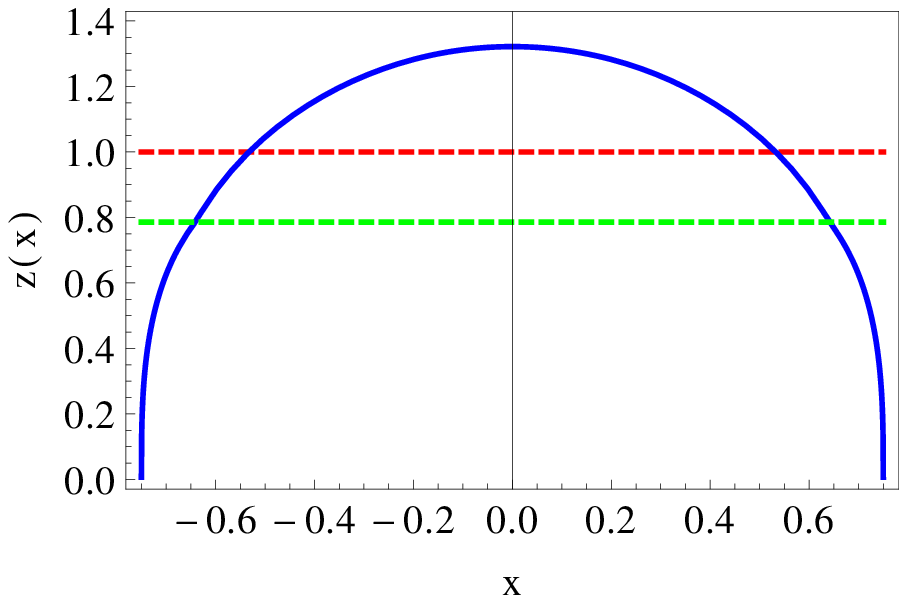}
\label{Wilsonprofile4} } 
\subfigure[$\gamma=0, t=0.802$]{
\includegraphics[width=0.3\columnwidth,height=0.2\columnwidth]{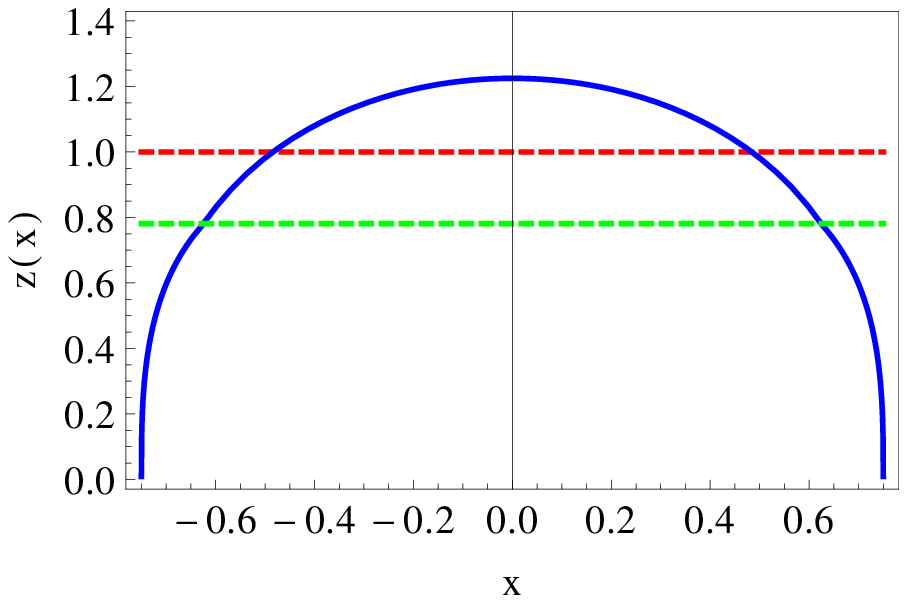}
\label{Wilsonprofile5} }
\subfigure[$\gamma=0.02, t=0.801$]{
\includegraphics[width=0.3\columnwidth,height=0.2\columnwidth]{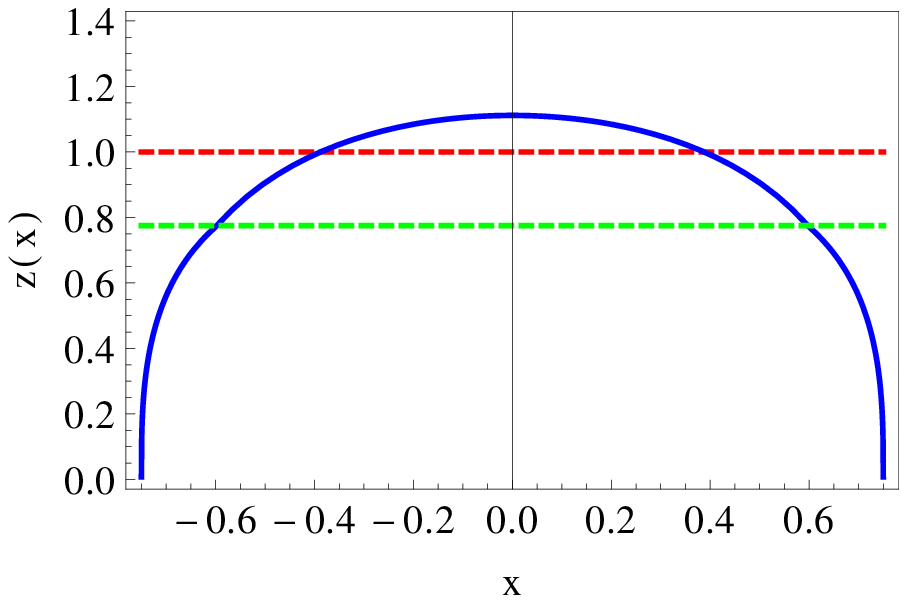}
\label{Wilsonprofile6} 
}
\subfigure[$\gamma=-0.01, t=1.001$]{
\includegraphics[width=0.3\columnwidth,height=0.2\columnwidth]{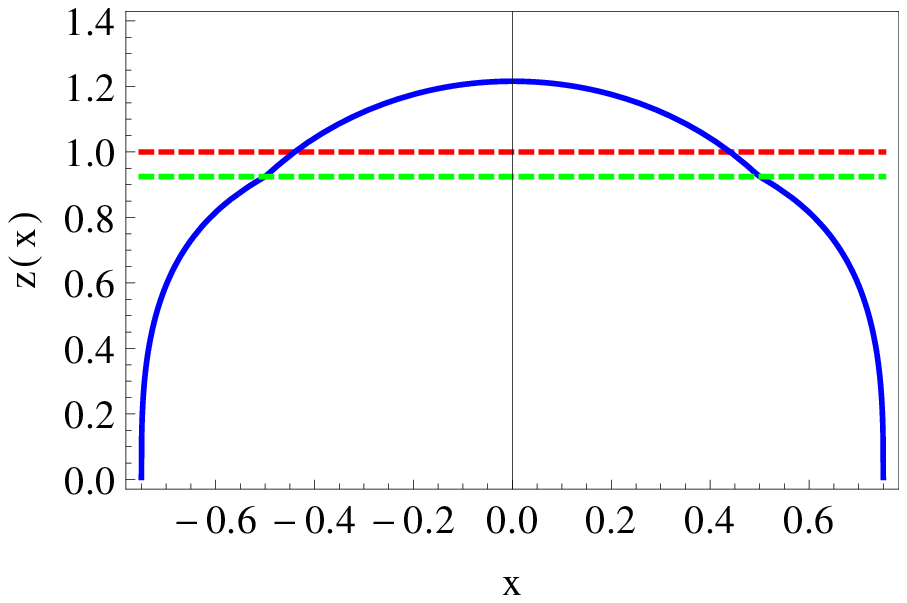}
\label{Wilsonprofile7} } 
\subfigure[$\gamma=0, t=1.003$]{
\includegraphics[width=0.3\columnwidth,height=0.2\columnwidth]{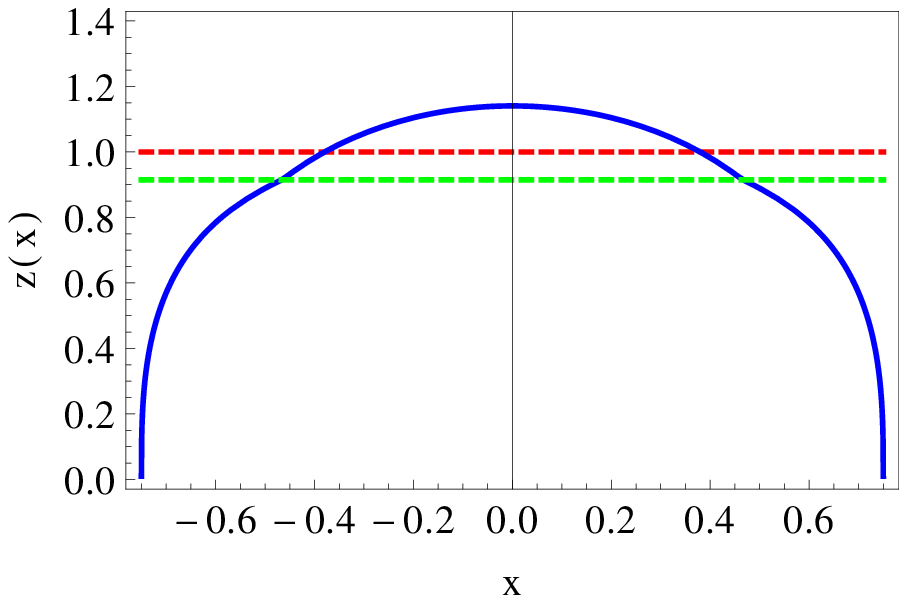}
\label{Wilsonprofile8} }
\subfigure[$\gamma=0.02, t=1.000$]{
\includegraphics[width=0.3\columnwidth,height=0.2\columnwidth]{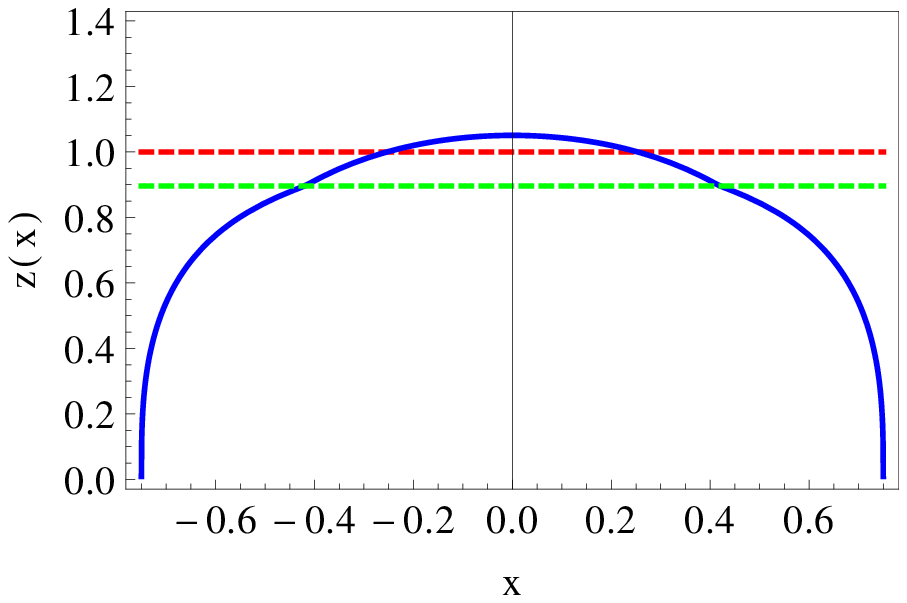}
\label{Wilsonprofile9} 
}
\subfigure[$\gamma=-0.01, t=1.348$]{
\includegraphics[width=0.3\columnwidth,height=0.2\columnwidth]{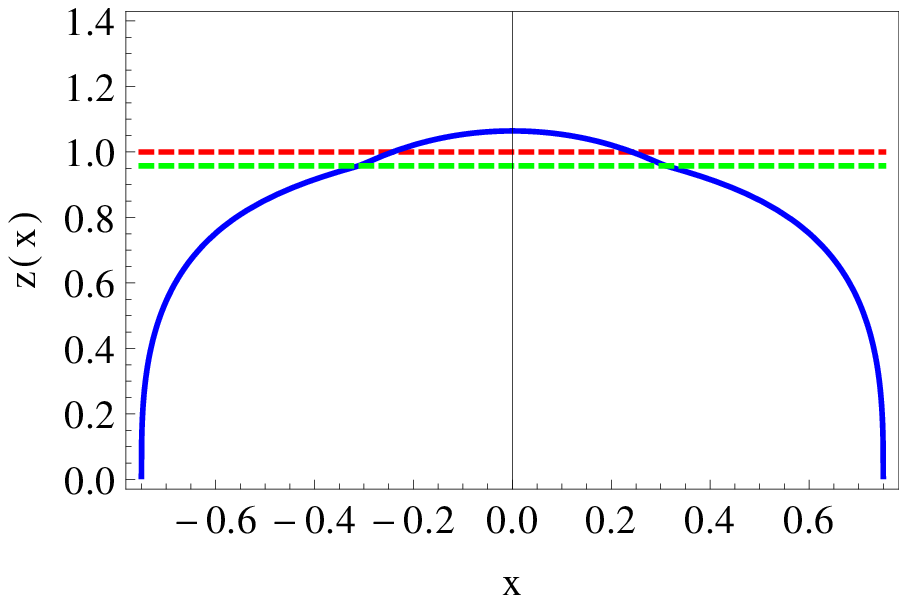}
\label{Wilsonprofile10} } 
\subfigure[$\gamma=0, t=1.354$]{
\includegraphics[width=0.3\columnwidth,height=0.2\columnwidth]{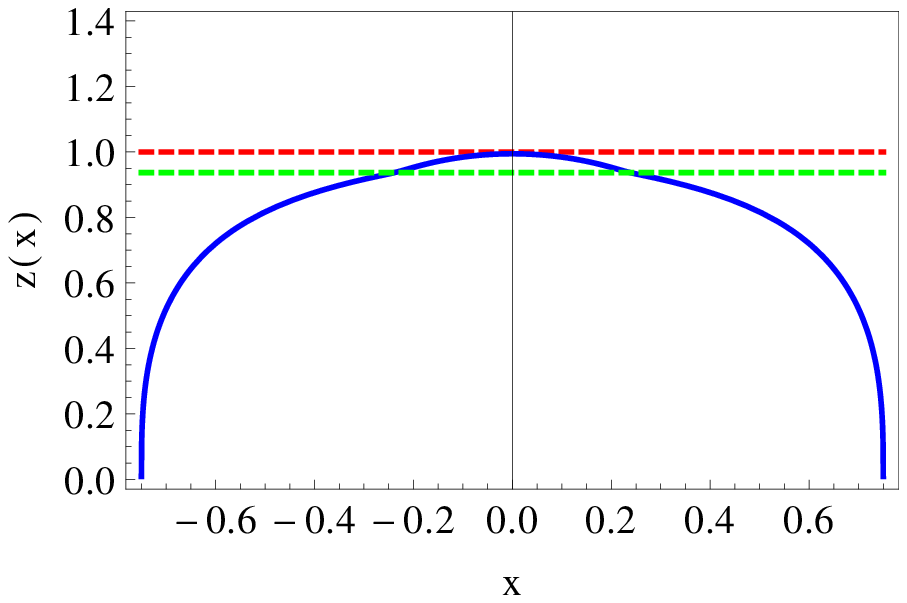}
\label{Wilsonprofile11} }
\subfigure[$\gamma=0.02, t=1.351$]{
\includegraphics[width=0.3\columnwidth,height=0.2\columnwidth]{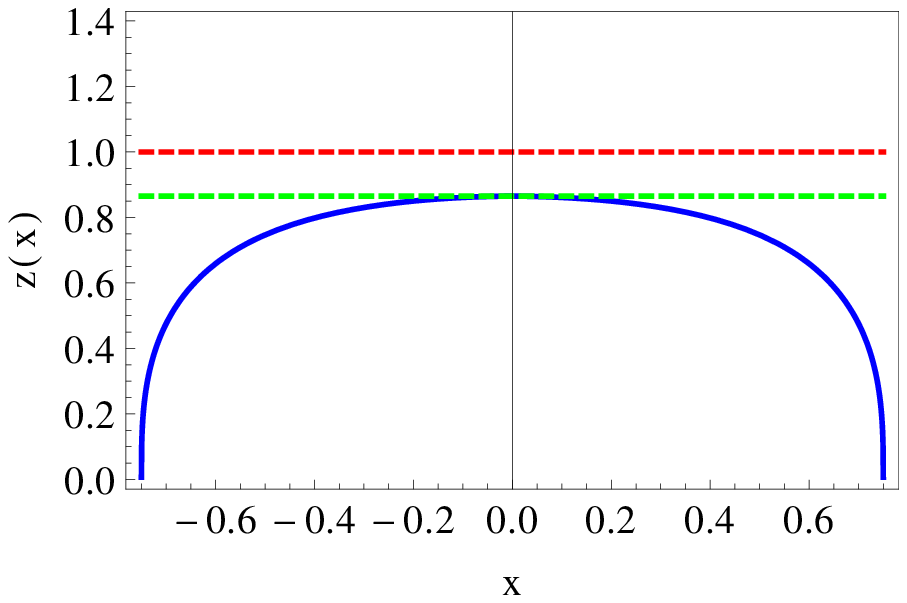}
\label{Wilsonprofile12} 
}
\caption{\small Time evolution of the minimal area surfaces and the position of the shell are shown for different values of $\gamma$ and 
fixed value of $Q=1$. The blue line denotes a cross-section of the minimal area surface at a particular boundary time, while 
the  green dashed line denotes the position of the shell at that particular time. The red dashed line at $z=1$ represents the 
horizon of the black brane to be formed at late time after the shell collapses. In all the cases the Wilson loop on the boundary
has length $l=1.5$ and width $R=2$. The left column corresponds to the time evolution of the minimal area surfaces
with $\gamma=-0.01$, the middle one corresponds to $\gamma=0$ and the right column represents the evolution for $\gamma=0.02$.}
\label{Wilsonprofile}
\end{figure}

\subsection{Wilson Loop and the Renormalized Minimal Area Surfaces}
In this subsection, we consider the time evolution of the minimal area surfaces by probing the Wilson loop into the bulk AdS space. Solving
the set of equations (\ref{WilsonVeomZeom}) with proper initial conditions as prescribed earlier, we generate a sequence of 
minimal area surface profiles at different times for different values of $\gamma$. This has been shown in 
figure \ref{Wilsonprofile} where we have choosen a rectangular Wilson loop of length $l=1.5$ and width $R=2$ on the boundary of the AdS
space and we have fixed the charge $Q=1$. The left column corresponds to the time evolution of the minimal area surfaces
for $\gamma=-0.01$, the middle one represents $\gamma=0$ and the right column corresponds to the time evolution with
$\gamma=0.02$. 

Note that as expected, the time evolution profiles are almost the same as the geodesic profiles in the previous 
subsection. As time elapses, the shell approaches towards the surface $z=1$ where the horizon of the black brane would be
formed at late time. Also note that there is a refraction of the minimal area surfaces on the shell when they cross 
the shell. As $\gamma$ becomes more negative, the minimal area surfaces penetrate more into the bulk. 
Now consider the three figures of the last row at a fixed boundary time $t \approx 1.125$. With $\gamma=0.02$ the surface 
propagates only in the black brane and does not cross the shell suggesting that the boundary system has been  
thermalized. But with $\gamma=0$ and with $\gamma=-0.01$ the minimal area surfaces are still crossing the shell and 
propagating into the pure AdS geometry and hence we do not yet have a thermal Wilson loop on the boundary.
Hence, as in the previous subsection, we expect that as we increase the Weyl coupling from a negative value to a positive value,
the thermalization in the boundary field theory would also be easier.  

\begin{figure}[t!]
\centering
\subfigure[$Q=1$]{
\includegraphics[scale=0.75]{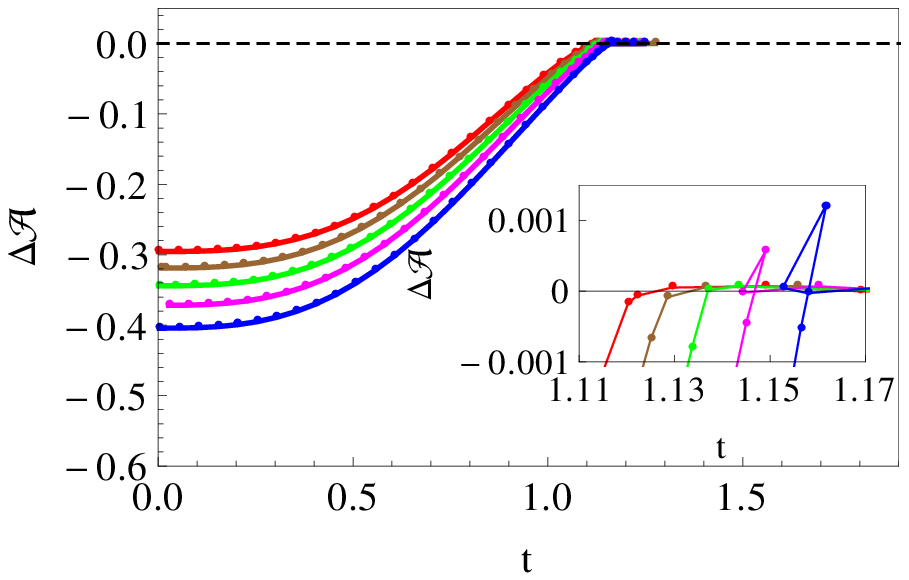}
\label{Area Plot Q 1} } 
\subfigure[$Q=\sqrt{2}$]{
\includegraphics[scale=0.75]{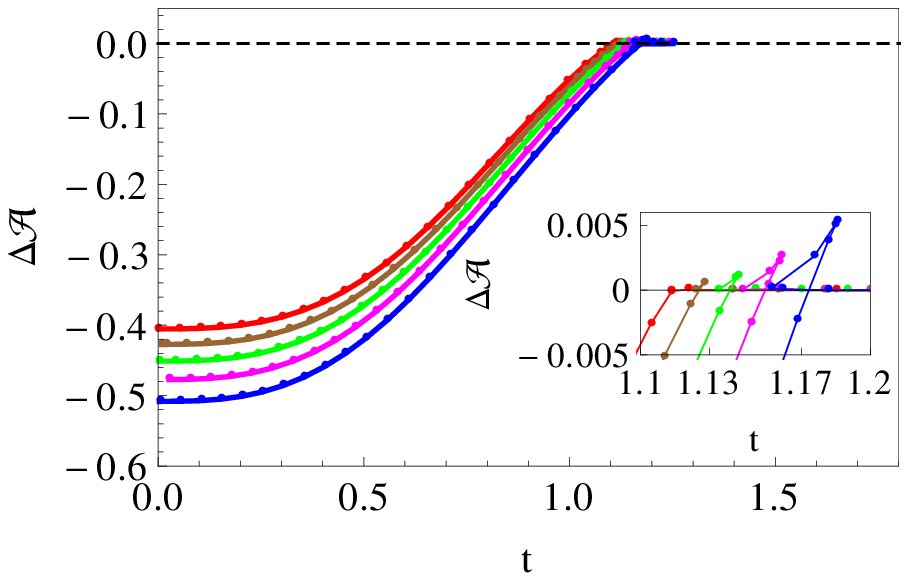}
\label{Area Plot Q 1.414} }
\caption{\small Time evolution of the renormalized minimal area for different values of $\gamma$ at fixed $Q$. The left one 
corresponds to the case $Q=1$ while the right one represents $Q=\sqrt{2}$. In each figure the curves with color
red, brown, green, magenta and blue correspond to $\gamma=0.02, 0.01, 0, -0.01$ and $-0.02$ respectively.}
\label{thermalization curves fixed Q probing minimal area}
 \end{figure}

Now we compute the dimensionless renormalized minimal area $\delta{A}$ and generate the thermalization curves to see whether
they are consistent with our expectation. This has been shown in figure \ref{thermalization curves fixed Q probing minimal area}
which describes the behaviour of the renormalized minimal area and hence the boundary Wilson loop with time for different 
values of the Weyl coupling constant $\gamma$ at fixed $Q$. We see a delay at the beginning of the thermalization time as in 
the case with two-point correlators. Starting with a negative value $\delta{A}$ increases with time and reaches zero at the
thermalization time.

In each figure, at a fixed value of $Q$, the thermalization time decreases as we increase the Weyl coupling from a negative
to a positive value which agrees with our result for the time evolution of the minimal area surfaces. The insets in each 
figure contain more details about the swallow-tail appearance. Notice that with $Q=1$ we have the swallow-tail appearance
only for negative values of $\gamma$. Remember that we did not get any swallow-tail emergence in the thermalization curve
for the two-point correlators with $Q=1$. Further, if we increase the charge to the extremal value $Q=\sqrt{2}$ even $\gamma=0$ and
$\gamma=0.01$ shows the swallow-tail behaviour before the thermalization. Hence, we conclude that the swallow-tail behaviour
is more prominent with negative values of $\gamma$ and large values of the charge parameter $Q$. Also for a given value of 
$\gamma$ and $Q$ probing the Wilson loop gives a clearer picture of the swallow-tail emergence than probing the two-point
correlator.

\begin{figure}[t!]
\centering
\subfigure[$\gamma=-0.01$]{
\includegraphics[scale=0.75]{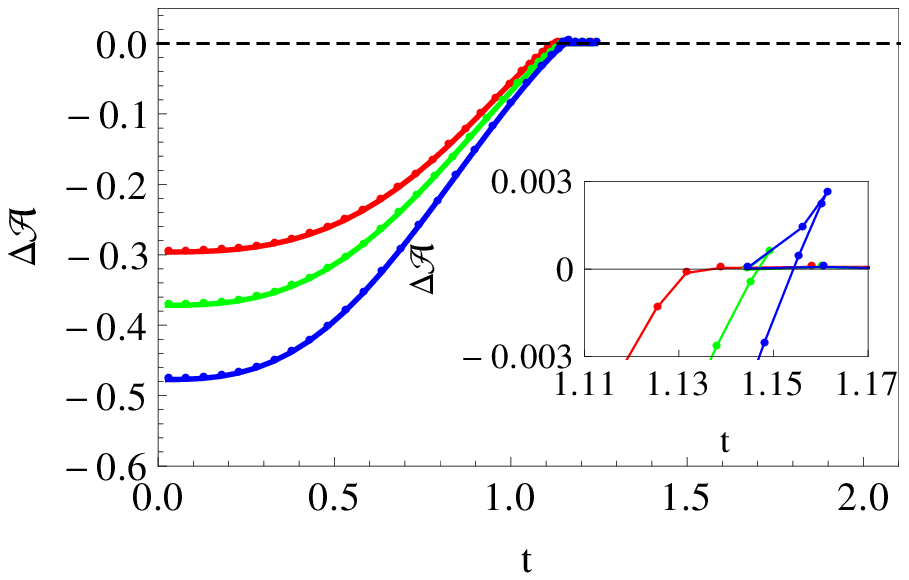}
\label{Area Plot Gamma -0.01} } 
\subfigure[$\gamma=0.02$]{
\includegraphics[scale=0.75]{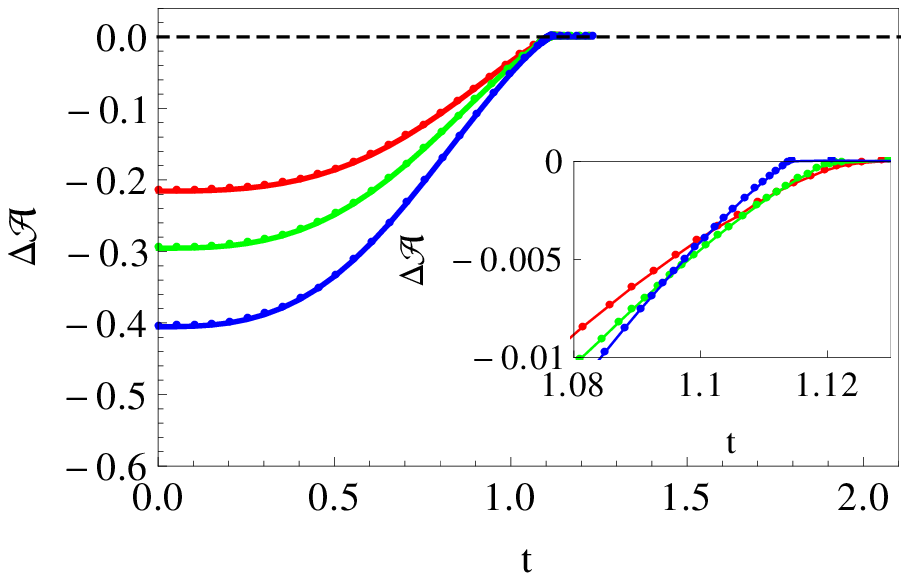}
\label{Area Plot Gamma 0.02} }
\caption{\small Time evolution of the renormalized minimal area for different values of $Q$ at fixed $\gamma$. The left one 
corresponds to the case $\gamma=-0.01$ while the right one represents $\gamma=0.02$. In each figure the curves with color
red, green and blue correspond to $Q=0.5, 1$ and $\sqrt{2}$ respectively.}
\label{thermalization curves fixed Gamma probing minimal area}
 \end{figure}

Figure \ref{thermalization curves fixed Gamma probing minimal area} shows the thermalization curves with $Q$ as a parameter
and keeping $\gamma$ fixed. Notice that for $\gamma=-0.01$, as $Q$ increases there is a delay in the thermalization time
as in the case with the two-point correlator. We also get a swallow-tail pattern for $Q=1$ and $Q=\sqrt{2}$. But as we 
fix $\gamma$ to a positive value, $\gamma=0.02$, we see again a negligible change in the thermalization time. But if we 
zoom the figure in the region just before the thermalization time, we see that $Q=\sqrt{2}$ takes the minimum time to
thermalize. Hence with $Q=\sqrt{2}$ even if the initial state of the dual field theory is much away from thermal equilibrium
than the states with $Q=1$ and $0.5$, but a shorter delay time at the beginning of the thermalization and a faster growth
of $\delta \mathcal{A}$ than the other two make the thermalization time for $Q=\sqrt{2}$ lesser. Hence for high 
positive values of $\gamma$, it is the large values of the charge parameter, which make the thermalization faster, although
the situation is different with negative values of $\gamma$.  This behavior was also noticed in \cite{Kundu} where for small values 
of the boundary separation, the boundary field theory thermalizes faster with large values of the charge parameter, while for large
separation, the boundary theory thermalizes later with large values of the charge parameter. These authors argued that with fixed value of the 
charge parameter there may exist two different regimes in the boundary field theory : for small $l$ the boundary theory is in quantum regime and 
for large $l$ it lies in the classical regime. We have checked that for large value of the boundary separation, it is the smaller values of the 
charge parameter which makes the thermalization faster. But it is unclear whether there exists a notion of classical and quantum regime, since
the system is out of equilibrium. Notice that there is no appearance of the swallow-tail pattern with 
larger positive values of $\gamma$ even with the extremal charge (we are assuming here that our perturbative analysis is valid for
such values of $\gamma$).

\begin{figure}[t!]
\centering
\subfigure[$Q=1$]{
\includegraphics[scale=0.75]{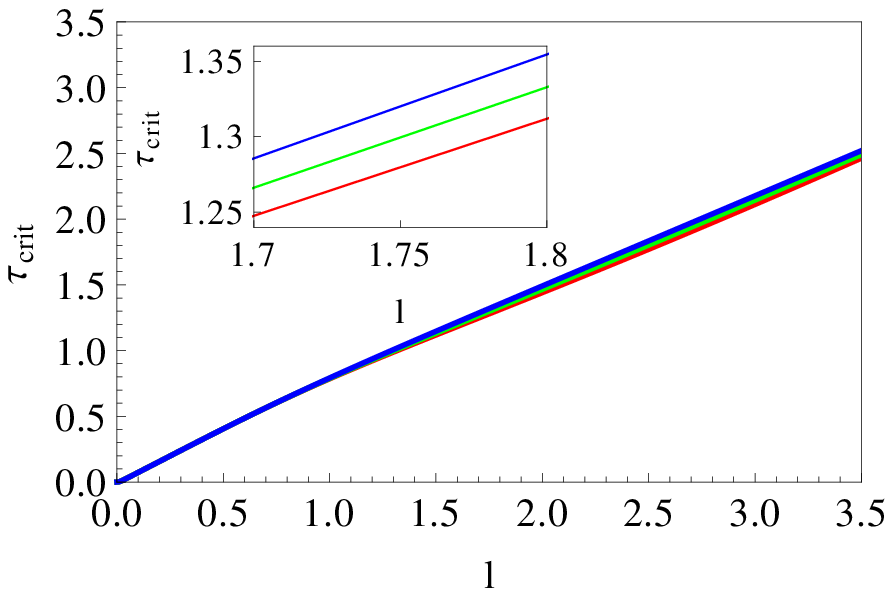}
\label{WL l vs tau Q1} } 
\subfigure[$Q=\sqrt{2}$]{
\includegraphics[scale=0.75]{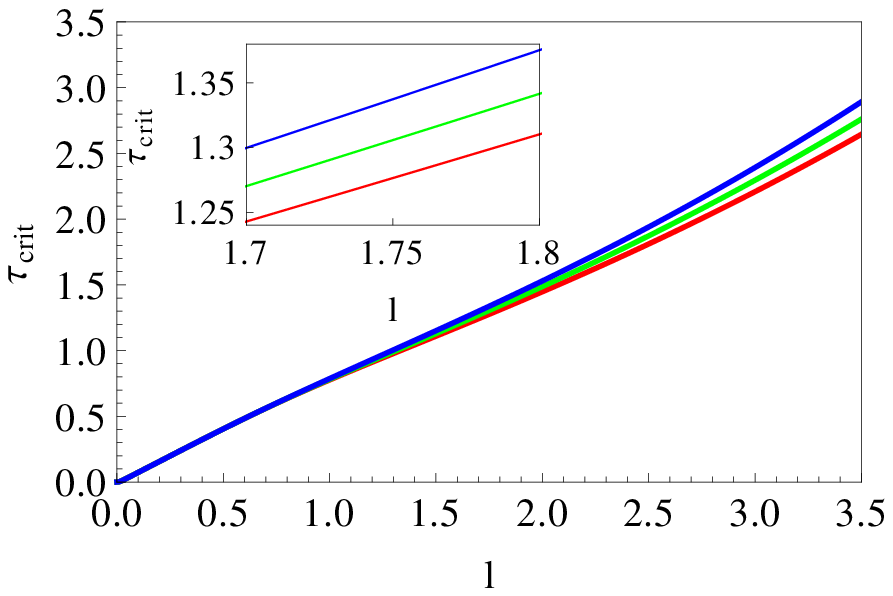}
\label{WL l vs tau Q1.414} }
\caption{\small $\tau_{crit}$ as a function of $l$ for different values of $\gamma$ at fixed $Q$. The left one 
corresponds to the case $Q=1$ while the right one represents $Q=\sqrt{2}$. The curves with color red, green and blue 
correspond to $\gamma=0.02$, $0$ and $-0.02$ respectively.}
\label{l vs tau curves fixed Q probing WL}
\end{figure}

Now we plot the critical time $\tau_{crit}$ as a function of the length of the rectangular Wilson loop $l$ while we fix 
the Width $R=2$. Figure \ref{l vs tau curves fixed Q probing WL} shows that for very small values of $l$, $\tau_{crit}$ is
almost independent of $\gamma$. The thermalization time no longer behaves as $\tau_{crit}\sim {l\over 2}$ in this case.
It is clear that as we increase $\gamma$, $\tau_{crit}$ decreases and this is more prominent when $Q$ is sufficiently large.

\section{Discussions and Conclusions}

In this section we summarize the main results of this paper. We have considered five dimensional AdS gravity 
coupled to a $U(1)$ gauge field by a combination of two and four derivative interactions, where, the four derivative interaction couples two powers 
of the Maxwell field to the bulk Weyl tensor. We call the coefficient of this four derivative interaction as the Weyl coupling constant $(\gamma)$ and have 
studied how this coupling along with the chemical potential affect thermalization in the dual gauge theory. This gives an extra control parameter that 
might be important to construct models of thermalization in realistic situations. For this purpose, first  
we constructed the black brane metric which solves the Einstein and Maxwell equations up to linear order in $\gamma$, since an exact analytical solution 
seems to be intractable. Then we constructed the Vaidya-like dynamical black brane metric which represents a pure AdS space at early times and a 
Weyl-corrected black brane at late times. We used two non-local observables on the boundary field theory to study the thermalization: equal time two-point 
correlation functions and the expectation value of the Wilson loop operator. Using the gauge/gravity duality, these two observables were identified to be the dual of
geometric quantities in the bulk gravity: the geodesic length and the minimal area surface, respectively, extending into the bulk from the 
boundary. These were then used to compute several physical quantities associated with the thermalization of the strongly coupled boundary theory. 

Broadly, we have presented all necessary physical details of thermalization in Weyl corrected five dimensional AdS gravity duals. Namely, we have analyzed the 
time evolution of geodesics and the time dependence of the geodesic length for two point correlators. Correspondingly, we have studied the 
time dependence of minimal area surfaces for rectangular Wilson loops. For both these cases, we analyzed the thermalization time scale.
The subtle interplay between the Weyl constant and the chemical potential has been elaborated upon, and we have seen that the
thermalization times for strongly coupled field theories may increase or decrease depending on the relative values of the two. An outcome of
our analysis was the appearance of a swallow-tail behaviour in the  thermalization curve, whose onset is affected by the Weyl coupling, and we have seen evidence 
that this might indicate distinct physical possibilities relating to different scales in the problem, assuming that the saddle point approximation in computing the 
thermalization time continues to be valid. 

One can also use another non-local observable, the entanglement entropy, to probe the thermalization in the boundary field theory. Using the 
gauge/gravity duality, it would be dual to the minimal volume extending into the bulk. In \cite {Bala}, it was shown that it is the entanglement 
entropy which sets the relevant time scale in the problem, since it thermalizes the last. In the present case we found it difficult to generate the
thermalization curves for holographic entanglement entropy, since the numerical computations there need a very high working precision. Let us elaborate
on this in some details. In Appendix C, we have provided the setup for calculating the Holographic entanglement entropy in the framework of Weyl 
corrected gravity (up to first order in $\gamma$). 

In principle, we can compute the holographic entanglement entropy numerically following exactly same procedure as we did 
for the two-point correlator and the Wilson loop. However, we find that this requires an enhancement of the working precision in MATHEMATICA
by a large amount in order for the numerical values to be reliable. For example, as we have already mentioned in section $5$, to produce the thermalization curves for 
the two-point correlator we have to first solve the pair of equations (\ref{geodesicVeomZeom}) subject to the initial conditions of 
(\ref{initial condition1}) for a fixed value of $\gamma$, $Q$ 
and $M$. To extract the boundary time $t$, we fix the value of $v_*$ and tune the value of $z_*$ until we get $z=z_0=0.01$ (as chosen in section $5$)
at the end point of the geodesic. Here, we adjust the value of $z_*$ in such a way that we get $z=z_0=0.01$ with 
${l\over 2} \in (1.499999, 1.500001)$, i.e., we have a tolerance of $10^{-6}$ determining the length of the boundary separation. With this tolerance,
we produce sensible results for the time evolution of the two-point correlators.

However, while dealing with the entanglement entropy, we have checked that a tolerance of $10^{-6}$ does not give trustable numerical results (which can be
compared for example with the entanglement entropy after setting $\gamma=0$, where a tolerance of $10^{-6}$ produces existing results in the literature). 
In fact, due to the complexity of the equation, we have to increase the tolerance upto $10^{-15}$ to rely on numerical values produced. 
Tuning the value of $z_*$ with this amount of tolerance is a daunting task. Although we have performed a limited analysis in this regard, 
and seen indication that the behaviour of the entanglement entropy
produces qualitatively similar results as from those of the other probes, a complete analysis seems rather tedious. 
So, although the procedure to evaluate the entanglement entropy is 
similar to the computation of the other two non-local probes, we did not establish in details the numerical values of the entanglement entropy here.

As pointed out in section $2$ (see discussion following (\ref{A ansatz})), our numerical analysis of section $5$ has a possible caveat. Namely, we have
chosen certain small values of the Weyl coupling $\gamma$ in the absence of a controlled perturbative expansion in powers of the same. 
Indeed, while a second order $\gamma$-corrected metric can be obtained, analysis of thermalization becomes difficult due
to the complexity of the equations. However, we have checked that in the present analysis, smaller numerical values of $\gamma$ (than those chosen here) 
does not change the qualitative aspects of our analysis, indicating that our numerical results up to ${\mathcal O}(\gamma)$ are trustable for the values 
of the coupling used in this paper. However we admit that this is a drawback of our numerical analysis. 

Arguably, our analysis has been limited to the fact that we have used a bottom-up phenomenological approach. Although understanding thermalization with 
the most general four derivative action in five dimensional AdS gravity might be difficult, it should be interesting to see if progress can be made on 
this.

\begin{center}
{\bf Acknowledgements}
\end{center}
AD and SM would like to thank Sayantani Bhattacharya for helpful discussions. AD also wishes to thank Damian Galante and Arnab Kundu 
for helpful email correspondence. SM would like to thank Joydeep Chakrabortty and Indian Institute of
Technology, Kanpur for hospitality and financial support. The work of SM is supported by project No. SPO/DST/PHY/20130115.

\appendix
\renewcommand{\theequation}{A.\arabic{equation}}
\setcounter{equation}{0} 
\section{Energy-Momentum Tensor}
For completeness here we briefly discuss how to derive the bulk energy-momentum tensor making use of the Palatini
identities. Using (\ref{WeylFF}), we can write down the action (\ref{action}) in the following form,
\begin{eqnarray}
&&\textit{S} = \frac{1}{16 \pi G_5} \int \mathrm{d^5}x \sqrt{-g} \ \ \bigl[R+\frac{12}{L^{2}}
-\frac{1}{4}\textit{F}_{\mu\nu}\textit{F}^{\mu\nu}+\gamma L^2\bigl(\textit{R}_{\mu \nu \rho \lambda}
\textit{F}^{\mu\nu}\textit{F}^{\rho\lambda}-{4\over 3}R_{\mu\nu}\textit{F}^{\mu}\hspace{0.1mm} _{\rho}\textit{F}^{\nu\rho}
\nonumber\\ 
&&+{1\over 6}R F_{\mu\nu}F^{\mu\nu}\bigr)\bigr] .
\label{appendix action}
\end{eqnarray}
Because of the non-zero Weyl coupling constant $\gamma$ the energy-momentum tensor would get a finite contribution from
the variation of the corresponding part in the action,
\begin{eqnarray}
\textit{S}_{Weyl} = \frac{1}{16 \pi G_5} \int \mathrm{d^5}x \sqrt{-g} \ \  \gamma L^2\bigl(\textit{R}_{\mu \nu \rho \lambda}
\textit{F}^{\mu\nu}\textit{F}^{\rho\lambda}-{4\over 3}R_{\mu\nu}\textit{F}^{\mu}\hspace{0.1mm} _{\rho}\textit{F}^{\nu\rho} 
+{1\over 6}R F_{\mu\nu}F^{\mu\nu}\bigr) \, . \ \
\label{appendix Weyl action}
\end{eqnarray}
The variation of which gives,
\begin{eqnarray}
&& \delta \textit{S}_{Weyl} = \gamma L^2 \int \mathrm{d^5}x \ \ \delta\bigl(\sqrt{-g}\bigr) \ \ \frac{1}{16 \pi G_5} \bigl[\textit{R}_{\mu \nu \rho \lambda}
\textit{F}^{\mu\nu}\textit{F}^{\rho\lambda}-{4\over 3}R_{\mu\nu}\textit{F}^{\mu}\hspace{0.1mm} _{\rho}\textit{F}^{\nu\rho} 
+{1\over 6}R F_{\mu\nu}F^{\mu\nu}\bigr]\nonumber \\
&& +\gamma L^2 \int \mathrm{d^5}x \sqrt{-g} \ \ \frac{1}{16 \pi G_5}\bigl[^{\mu\nu}F^{\rho\lambda}\delta R_{\mu\nu\rho\lambda}+
R_{\mu\nu\rho\lambda}\delta\bigl(F^{\mu\nu}F^{\rho\lambda}\bigr)-{4\over 3}\textit{F}^{\mu}\hspace{0.1mm}
_{\rho}\textit{F}^{\nu\rho}\delta R_{\mu\nu} \nonumber \\
&&-{4\over 3}R_{\mu\nu}\delta\bigl(\textit{F}^{\mu}\hspace{0.1mm} _{\rho}\textit{F}^{\nu\rho}\bigr)
++{1\over 6} F_{\mu\nu}F^{\mu\nu} \delta R +{1\over 6}R \  \delta \bigl(F_{\mu\nu}F^{\mu\nu}\bigr)\bigr] .
\label{appendix Weyl action variation}
\end{eqnarray}
Now to evaluate the variations of $R_{\mu\nu\rho\lambda}$, $R_{\mu\nu}$ and $R$, we use the Paalatini identities,
\begin{eqnarray}
 \delta R^{\rho}\hspace{0.1mm}_{\sigma\mu\nu}=\nabla _{\mu}(\delta \Gamma ^{\rho}_{\nu\sigma})-
 \nabla _{\nu}(\delta \Gamma ^{\rho}_{\mu\sigma}) \,,\nonumber \\
 \delta R_{\mu\nu}=\nabla _{\rho}(\delta \Gamma ^{\rho}_{\nu\mu})-
 \nabla _{\nu}(\delta \Gamma ^{\rho}_{\rho\mu}) .
 \label{Palatini identities}
\end{eqnarray}
Also we use the fact that the variation of the Christoffel symbol is a tensor given by
\begin{eqnarray}
 \delta \Gamma ^{\rho}_{\mu\nu}={1\over 2}g^{\rho\lambda}\bigl(\nabla_{\mu}\delta g_{\lambda\nu}+
 \nabla_{\nu}\delta g_{\lambda\mu}-\nabla_{\lambda}\delta g_{\mu\nu}\bigr) .
\end{eqnarray}
Then after few steps one can derive the following expressions,
\begin{eqnarray}
 &&\delta R_{\mu\nu\rho\lambda}=-g_{\mu\alpha}R_{\beta\nu\rho\lambda}\delta g^{\alpha\beta}+{1\over 2}\bigl(
 \nabla_{\rho}\nabla_{\lambda}\delta g_{\mu\nu}
 -\nabla_{\lambda}\nabla_{\rho}\delta g_{\mu\nu} +\nabla_{\rho}\nabla_{\nu}\delta g_{\mu\lambda}\nonumber \\
  &&-\nabla_{\lambda}\nabla_{\nu}\delta g_{\mu\rho}-\nabla_{\rho}\nabla_{\mu}\delta g_{\lambda\nu}+
  \nabla_{\lambda}\nabla_{\mu}\delta g_{\rho\nu}\bigr) \,,\nonumber\\
 &&\delta R_{\mu\nu}={1\over 2}g^{\rho\lambda} \bigl(\nabla_{\rho}\nabla_{\nu}\delta g_{\lambda\mu}
 +\nabla_{\rho}\nabla_{\mu}\delta g_{\lambda\nu}-\nabla_{\rho}\nabla_{\lambda}\delta g_{\nu\mu}
 -\nabla_{\nu}\nabla_{\rho}\delta g_{\lambda\mu}-\nabla_{\nu}\nabla_{\mu}\delta g_{\lambda\rho}\nonumber\\
 &&+\nabla_{\nu}\nabla_{\lambda}\delta g_{\rho\mu}\bigr)\,,\nonumber\\
 &&\delta R= \bigl(R_{\alpha\beta}+g_{\alpha\beta}\Box-\nabla_{\alpha}\nabla_{\beta}\bigr)\delta g^{\alpha\beta} .
 \label{variation of Riemann Tensor}
\end{eqnarray}
Now having all the expressions for the variations of $R_{\mu\nu\rho\lambda}$, $R_{\mu\nu}$ and $R$, substituting them in 
(\ref{appendix Weyl action variation}) we can derive the expression for the energy-momentum tensor $T_{\mu\nu}$ given in 
(\ref{EnergyMomentumTensor}) and hence write down the Einstein equation.

\renewcommand{\theequation}{B.\arabic{equation}}
\setcounter{equation}{0}
\section{Solution with a different metric ansatz}
Following \cite{Myers} one can choose a metric ansatz different from the ansatz chosen in section 2, i.e., one can 
choose,
\begin{eqnarray}
ds^2&=&- \frac{r^2 f(r)}{L^2}  dt^2+\frac{L^2}{r^2 g(r)}dr^2
+\frac{r^2}{L^2}(dx^2+dy^2+d\eta^2)\,,\nonumber \\
A&=& (\phi(r),0,0,0,0)\,. 
\label{different metric ansatz}
\end{eqnarray}
Again we have to solve it up to linear order in $\gamma$ since an exact analytical solution for the metric seems to 
be imposible. We consider the same form as \cite{Myers} for $f(r)$, $g(r)$ and $\phi(r)$
\begin{eqnarray}
f(r)&=&f_0(r)[1+F(r)]\,,\nonumber\\
g(r) &=& f_0(r)[1+F(r)+G(r)]\,,\\
\phi(r) &=& \phi_0(r)+ \psi(r) .\ \nonumber
\label{different perturbation ansatz}
\end{eqnarray}
where $f_0(r)$ and $\phi_0(r)$ are the zeroth order solutions representing a Reissner Nordstr\"{o}m  black brane given by 
\begin{eqnarray}
 f_{0}(r) &=& 1-\frac{M L^2}{r^4}+\frac{Q^2 L^2}{r^6} \,,\nonumber\\
  \phi_{0}(r) &=& {L^3\over 2} q \bigl({1\over r_h^2}-{1\over r^2}\bigr) .
 \label{different zeroth order soln}
\end{eqnarray}
where $q=\left(*F\right)_{xy\eta}=2\sqrt{3}{Q\over L^3}$ represents the charge density and `$r_h$' denotes the position 
of the event horizon.

Again we solve the equations (\ref{EinsteinEOM}) and (\ref{MaxwellEOM}) to linear order in $\gamma$ as we did in 
section 2 and get the perturbations $F(r)$, $G(r)$ and $\psi(r)$ representing the $O(\gamma)$ corrections,
\begin{eqnarray}
  F(r)&=& {\gamma \over f_0(r)} \bigl(-k_2+{r_h^4 \over r^4}k_1+{2 L^2 Q^2\over r^6}k_4-{7L^4 Q^4\over r^{12}}+{8L^4 M Q^2\over r^{10}}
  -{16L^2 Q^2\over r^{6}}\bigr) \,,\nonumber\\
  G(r)&=&  \gamma \bigl(k_2-\frac{8 L^2 Q^2}{r^{6}}\bigr) \,,\\
   \psi(r)&=& \gamma \bigl(k_3-{\sqrt{3}Q\over r^2}k_4+{14\sqrt{3}L^2 Q^3\over r^8}-{8\sqrt{3}L^2 M \ Q\over r^6}\bigr) .\ \nonumber
   \label{different perturbations}
\end{eqnarray}
where $k_1$, $k_2$, $k_3$ and $k_4$ are dimensionless integration constants.
Imposing the same constraints on the above equations as in section 2, one can evaluate those integration constants and 
write down the final form of the metric perturbations,
\begin{eqnarray}
 F(r)&=& { \gamma \over f_0(r)} \bigl(-\frac{L^4 M^2}{r^4 r_h^4}+\frac{8L^4 M Q^2}{r^{10}}+\frac{10L^2 M}{r^4}-\frac{7
   L^4 Q^4}{r^{12}}-\frac{16L^2 Q^2}{r^6}-\frac{9 r_h^4}{r^4}\bigr) \,,\nonumber\\
  G(r)&=& - \gamma \frac{8 L^2 Q^2}{r^6} \,,\\
   \psi(r)&=& \gamma \bigl(-\frac{8 \sqrt{3} L^2 M Q}{r^6}+\frac{8 \sqrt{3}L^2 M Q}{r_h^6}+\frac{14 \sqrt{3}L^2 Q^3}{
   r^8}-\frac{14 \sqrt{3}L^2 Q^3}{r_h^8}\bigr) .\ \nonumber
   \label{different Final perturbations}
\end{eqnarray}
Now introducing $z={L^2\over r}$ and writing down this metric and gauge field in the Eddington-Finkelstein coordinate 
\begin{eqnarray}
 dv=dt-{dz\over \sqrt{f(z) g(z)}} \,,
 \label{different EFcoord}
\end{eqnarray}
we have,
\begin{eqnarray}
 ds^2&=&{L^2\over z^2}\bigl(-f(z) dv^2-2 \sqrt{{f(z)\over g(z)}} dv dz+dx^2+dy^2+d\eta^2\bigr)\,, \nonumber \\
A&=& \phi(z)\bigl(dv+{dz\over \sqrt{f(z)g(z)}}\bigr)\,. 
\label{defferent metricEF}
\end{eqnarray}
Again using a proper gauge, we set $A_z=0$ and the gauge field becomes $A=A_t dv=\phi(z)dv$.

Now if we consider the mass and charge parameter to depend on the advanced time coordinate $v$, the function $F$, $G$
and $\psi$ would also explicitly depend on $v$. So we need to introduce an external matter source to satisfy the Einstein
and Maxwell equation given by (\ref{EinsteinMaxwellVaidyaEOM}) where the external matter source would satisfy,  
  \begin{eqnarray}
  && T_{v v}^{(ext)}=-\frac{3}{2} {z^3\over L^{10}} \bigl(2 z^2 Q(v) Q'(v)-L^4 M'(v)\bigr) -\frac{3}{2}{z^3\over L^{10}}
  \gamma\bigl[-2 {L^6\over r_h^4}M(v) M'(v) \nonumber \\
   && +10 L^4 M'(v)-32 {z^6\over L^6} M(v) Q(v) Q'(v)-16 z^3 Q(v) Q''(v)+20 {z^8\over L^{10}} Q(v)^3 Q'(v) \nonumber \\
   && -16 z^3 Q'(v)^2+48 z^2 Q(v) Q'(v)\bigr] \,,\nonumber \\
   && T_{v z}^{(ext)}= -24 \gamma  {z^5\over L^{10}} Q(v) Q'(v) \,, \nonumber \\ [10pt]
   && T_{x x}^{(ext)}= T_{y y}^{(ext)}= T_{\eta\eta}^{(ext)}=-96 \gamma  
   {z^5\over L^{10}} Q(v) Q'(v).\nonumber \\
   \label{differentEinsteinVaidyaEOM}
\end{eqnarray}
\begin{eqnarray}
  J^{\lambda}_{(ext)}=-2 \sqrt{3} {z^5\over L^8} Q'(v) \bigl(1-4 \gamma  {z^6\over L^{10}} Q(v)^2\bigr)~\delta^{\lambda}_z.  
  \label{differentMaxwellVaidyaEOM}
  \end{eqnarray}
Now constructing a null shell of charged fluid with this energy-momentum tensor and checking the null energy conditions
with the mass and charge functions given in (\ref{mq}) we can write down a similar set of equations like
(\ref{geodesicVeomZeom}) and (\ref{WilsonVeomZeom}) for the geodesic and the minimal area surfaces. We noticed that
when the boundary time $t$ is small, we can solve the the two coupled equations and calculate $\delta \mathcal{L}$
and $\delta \mathcal{A}$. But for large value of $t$ the differential equations exhibit some form of stiffness and 
we could not get a stable solution for $z(x)$ and $v(x)$. So we could not get a complete thermalization curve 
both for the two-point correlator and the Wilson loop. Hence we switched to a different metric ansatz as explained in
section 2 and resolved this issue. With that metric ansatz the differential equations 
(\ref{geodesicVeomZeom}) and (\ref{WilsonVeomZeom}) did not exhibit any kind of stiffness problem. 
However, we have checked that, both the metric ansatzs give the same results (same up to five decimal places) for small value of $t$, as expected.

\renewcommand{\theequation}{C.\arabic{equation}}
\setcounter{equation}{0}
\section{Holographic Entanglement Entropy}
We can use entanglement entropy as another tool for probing the thermalization. For the sake of completeness, we discuss the basic features
of the same in this appendix. 
If our boundary system is divided into two subsystems $\mathcal{A}$ and its complement $\mathcal{B}$, the entanglement entropy of the subsystem 
$\mathcal{A}$ is defined as,
\begin{eqnarray}
 S_{\mathcal{A}}=-Tr_{\mathcal{A}}(\rho_{\mathcal{A}} \ln \rho_{\mathcal{A}}) .
 \label{EE}
\end{eqnarray}
where $\rho_{\mathcal{A}}$ is the reduced density matrix of $\mathcal{A}$, obtained by considering the trace over the degrees of freedom 
of $\mathcal{B}$, i.e., $\rho_{\mathcal{A}}=Tr_{\mathcal{B}}(\rho)$, where $\rho$ is the density matrix of the full quantum system. It is known that
direct calculation of entanglement entropy in a quantum field theory is difficult beyond $1+1$ dimensions. However, it becomes tractable if one uses the  
Ryu-Takayanagi formula \cite{RyuTakayanagi}. Using this formula, the holographic entanglement entropy of the subsystem $\mathcal{A}$ that
lives on the boundary of the AdS space is 
 \begin{eqnarray}
S_{\mathcal{A}}={\mbox{Area} (\Gamma _{\mathcal{A}})\over 4G_N} .
\end{eqnarray}
Here, $G_N$ is the Newton's constant of the bulk theory and $\Gamma _{\mathcal{A}}$ is a codimension-2  minimal-area 
hypersurface that extends into the AdS bulk, and it shares the same boundary $\partial{\mathcal{A}}$ as that of the subsystem $\mathcal{A}$.
However, as is known, this formula is only applicable to static backgrounds in the absence of higher derivative terms in the bulk action. 
In presence of such higher derivative corrections, the Ryu-Takayanagi conjecture no longer holds. In \cite{Dong}, a formula for holographic
entanglement entropy for a general higher derivative gravity theory was derived. This was shown to consist of the Wald entropy as the leading term,
with subleading corrections due to the extrinsic curvature. However, 
one can explicitly check that, the extra four derivative interaction term in our Lagrangian, 
$C_{\mu\nu\rho\lambda}F^{\mu\nu}F^{\rho\lambda}$, does not give rise to any subleading term in the covariant expression of the holographic entanglement 
entropy given in \cite{Dong}. Hence, for our purposes, we can use the expression of holographic entanglement entropy as 
\begin{eqnarray}
 S_{\rm EE} = -2\pi \int d^3 x \sqrt h \frac{\partial \tilde{\mathcal{L}}}{\partial R_{\mu\nu\rho\lambda}} \varepsilon_{\mu\nu} \varepsilon_{\rho\lambda} .
 \label{HEEWald}
\end{eqnarray}
where, $\tilde{\mathcal{L}}$ is the Lagrangian obtained from (\ref{action}) and $\varepsilon_{\mu\nu}$ is the binormal Killing vector, normalised as 
$\varepsilon_{\mu\nu} \varepsilon_{\mu\nu}=-2$. Thus we have
\begin{eqnarray}
S_{\rm EE}  &=& {1\over 4 G} \int_{-l/2}^{l/2} dx\left[{L^3 \over z(x)^3}
\sqrt{1-2 e^{-\chi(v,z)}v'(x)z'(x) -f(v,z) e^{-2\chi(v,z)} v'(x)^2}\times\right.\nonumber\\
~~~~~~~~~&~&\left.\Bigl(1-\gamma{z(x)^4\over L^2} \partial_z \phi_0(v,z)^2 \Bigr)\right] \, 
\label{minimal volume}
\end{eqnarray}
Like the previous cases, we will again have a conserved quantity, given by 
\begin{eqnarray}
 \tilde{ \mathcal{H}_2}  =  \frac{\Bigl({1\over z(x)^3}-\gamma {z(x)
 \over L^2}\partial_z \phi_0(v,z)^2\Bigr)}{\sqrt{1-2 e^{-\chi(v,z)}v'(x)z'(x) -f(v,z) e^{-2\chi(v,z)} v'(x)^2} } \, 
 \label{Hamiltonian 3}
\end{eqnarray}
Since the turning point of the codimension-2 hypersurface is at $x=0$ because of the symmetry of the problem, we can impose 
the initial conditions,
\begin{eqnarray}
v(0)=v_* ,\ \ \,   z(0)=z_* ,\ \ \,  v'(0)=0 ,\ \ \,  z'(0) = 0 .\ 
\label{initial condition3}
\end{eqnarray}
Using these initial conditions, (\ref{Hamiltonian 3}) simplifies to
\begin{eqnarray}
 \sqrt{1-2 e^{-\chi(v,z)}v'(x)z'(x) -f(v,z) e^{-2\chi(v,z)} v'(x)^2}={z_*^3\over z(x)^3}{\Bigl(1-\gamma {z(x)^4
 \over L^2}\partial_z \phi_0(v,z)^2\Bigr)\over \Bigl(1-12 \gamma {z_*^6\over L^{10}}q(v_*)^2\Bigr)}
\label{Conservation3}
\end{eqnarray}
We now minimize the functional given in (\ref{minimal volume}) and get the equations for $v(x)$ and $z(x)$ :
\begin{eqnarray}
&& f(v,z)v''(x)+v'(x)z'(x)\partial_z f(v,z)+{v'(x)^2\over 2} \partial_{v}f(v,z)-2v'(x)z'(x)f(v,z)\partial_{z}\chi(v,z)\nonumber \\
&&-f(v,z)v'(x)^2\partial_{v}\chi(v,z)+e^{\chi(v,z)}\bigl(z''(x)-z'(x)^2 \partial_{z}\chi(v,z)\bigr)\nonumber\\
&&-2\gamma{z(x)\over L^2 \mathcal{H}_2^2}\Bigl({1\over z(x)^3}-\gamma {z(x)\over L^2}\partial_z \phi_0(v,z)^2\Bigr)
e^{2\chi(v,z)}\partial_z \phi_0(v,z)\partial_v\partial_z \phi_0(v,z)=0 \ ,\nonumber \\
&& v''(x) e^{\chi(v,z)}-v'(x)^2\partial_{v}\chi(v,z)e^{\chi(v,z)}-{v'(x)^2\over 2}\partial_z f(v,z)+f(v,z)v'(x)^2\partial_z \chi(v,z)\nonumber\\
&&-{3\over \mathcal{H}_2^2 z(x)^7}e^{2\chi(v,z)}\Bigl(1-{\gamma{z(x)^4\over L^2}\partial_z\phi_0(v,z)^2}\Bigr)^2
-{\gamma \over L^2 z(x)^6 \mathcal{H}_2^2}\Bigl(1-\gamma {z(x)^4\over L^2}\partial_z \phi_0(v,z)^2\Bigr)\nonumber\\
&&e^{2\chi(v,z)}\Bigl(4z(x)^3\partial_z\phi_0(v,z)^2+2 z(x)^4\partial_z \phi_0(v,z) \partial_z^2 \phi_0(v,z)\Bigr)=0 \ .
\label{HEEVeomZeom}
\end{eqnarray}
We have to solve these equations numerically the same way we did for the the two-point correlator and the Wilson loop. 
Using the conservation equation (\ref{Conservation3}), we express the extremized area as,
\begin{eqnarray}
\mathcal{S}(\Gamma _{\mathcal{A}})  =  {2 \over 4 G}  \int_{0}^{l/2} dx \frac{z_*^3 L^3}{z(x)^6}
{\Bigl(1-12 \gamma {z^6\over L^{10}}q(v)^2\Bigr)\over \Bigl(1-12 \gamma {z_*^6\over L^{10}}q(v_*)^2\Bigr)}\equiv \mathcal{S} (l, t) \, .
\label{minimal volume2}
\end{eqnarray}
Equation (\ref{minimal volume2}) diverges because of the contribution from $z=0$.  So we again define 
a renormalized entanglement entropy by subtracting the divergent part as
\begin{eqnarray}
\mathcal{S}_{ren}(l,t) = \mathcal{S}(l,t) -  {1\over 2 z_0^2} \ .
\label{renormalized minimal volume}
\end{eqnarray}
where, $z_0$ is the UV cut-off of the theory. It turns out that numerical computation of the entanglement entropy becomes 
somewhat difficult in Weyl corrected gravity, even at first order in the Weyl coupling. This is explained in details in the last section of this paper.

\end{document}